\documentclass{article}
\usepackage[utf8]{inputenc}
\usepackage{authblk}

\usepackage{amsmath}
\usepackage[]{graphicx}
\usepackage[left=1cm,right=1cm, top=2cm,bottom=2cm,bindingoffset=0cm]{geometry}
\usepackage[usenames]{color}
\usepackage{colortbl}
\usepackage{subcaption}
\usepackage{physics}
\newcommand{\hp}{\hat{p}}
\newcommand{\h}{\hat}
\usepackage{amssymb}
\usepackage{bm}
\usepackage{xcolor}
\usepackage{hyperref}
\usepackage{comment}
\newcommand{\expv}[1]{\left\langle #1 \right\rangle}
\usepackage{indentfirst}
\usepackage{cleveref}
\usepackage{bigints}
\usepackage[normalem]{ulem}
\usepackage{leftindex}
\usepackage{natbib}
\bibpunct{[}{]}{,}{n}{,}{,}

\begin{document}

\title{Emission of twisted photons by a Dirac electron \\ in a strong magnetic field}
\author{I.\,Pavlov\thanks{ilya.pavlov@metalab.ifmo.ru} } 
\author{D.\,Karlovets\thanks{dmitry.karlovets@metalab.ifmo.ru}}
\affil{{\small School of Physics and Engineering, ITMO University, \\ 
197101 St.\,Petersburg, Russia}}

\date{\today}

\maketitle

\begin{abstract}
    We study spontaneous emission of a photon during the transitions between relativistic Landau states of an electron in a constant magnetic field that can reach the Schwinger value of $H_c = 4.4 \times 10^9$ T. In contrast to the conventional method in which detection of both the final electron and the photon is implied in a certain basis, here we 
    derive the photon state as it evolves from the process itself.
    It is shown that the emitted photon state represents a twisted Bessel beam propagating along the field axis with a total angular momentum (TAM) projection onto this axis $\ell-\ell'$ 
    where $\ell$ and $\ell'$ are the TAM of the initial electron and of the final one, respectively. Thus, the majority of the emitted photons turn out to be twisted with $\ell-\ell' \gtrsim 1$, even when the magnetic field reaches the critical value of $H\sim H_c$. The transitions without a change of the electron angular momentum, $\ell'=\ell$, are possible, yet much less probable. 
    We also compare our findings with those for a spinless charged particle and demonstrate their good agreement for the transitions without change of the electron spin projection even in the critical fields, while the spin-flip transitions are generally suppressed. In addition, we argue that whereas the ambiguous choice of an electron spin operator affects the differential probability of emission, this problem can partially be circumvented for the photon evolved state because it is the electron TAM rather than the spin alone that defines the TAM of the emitted twisted photon.

\end{abstract}

\section{Introduction}
According to classical electrodynamics, an electron moving along a helical path in an external electromagnetic field -- such as a magnetic field, that of a helical undulator, etc. -- emits electromagnetic waves possessing helical phase structure and carrying orbital angular momentum (OAM) \cite{katoh2017helical, katoh2017angular, epp2019angular, epp2022angular}. In quantum theory, the emitted field comprises the so-called twisted photons \cite{allen1992orbital, andrews2012angular, torres2011twisted, SerboUFN}. The quantum description of this phenomenon is complicated by the entanglement between the photon and the electron, which is rarely given attention in theoretical works \cite{EPJC}. Furthermore, the conventional quantum theory of synchrotron radiation \cite{ST,bordovitsyn1999synchrotron} deals with the so-called detected states of the final particles silently implying that both the electron and the photon are detected, and the photon detector state is usually chosen to be a plane wave. More recently, the photon detector state was chosen to be a Bessel beam with a definite projection of the total angular momentum (TAM) onto the magnetic field axis \cite{Kruining2019, Maruyama, Zhang2020}.  However, even more general problem would be to derive a quantum state of the photon as it has evolved from the emission process itself, without specifying the detector. In particular, such calculations in the scalar QED show that a spinless charged particle in the magnetic field indeed emits a Bessel beam \cite{Karlovets2023}, the parameters of which can be derived rather than postulated. Here, we present an analogous evolved state analysis for a spin-$\frac{1}{2}$ Dirac electron and compare it to the scalar case.

The magnetic field-induced Landau levels of charged particles are integral to the quantum theory of synchrotron radiation, which is essential in describing astrophysical objects like neutron stars, operating modern storage rings, and producing spin-polarized electron beams. These states also have significant implications in solid-state physics, particularly in graphene \cite{Sadovsky, Li, Yang},  including quantum Hall effect \cite{LLsplitting}, in diamagnetism of metals, in plasma \cite{Shah}, and in the quantum dynamics of charged particles in Penning traps \cite{Larson_1988, Kluge_2003}. Therefore, the fact that it is the twisted photons that are mostly emitted by electrons in magnetic fields -- both in the astrophysical and terrestrial environments -- could put a new twist on several fields. In this paper, we study single photon emission in the first order of the perturbation theory by an electron moving along the field lines of a constant and homogeneous magnetic field $\bm{H} = \{0, 0, H\}$ of an arbitrary strength $H > 0$ \cite{ST, bordovitsyn1999synchrotron}. The magnetic field is taken into account exactly in the calculations and the electron is described with the relativistic Landau states \cite{ST, bordovitsyn1999synchrotron, bagrov2014dirac}. 

One of our main findings is that the evolved state of the emitted photon is always twisted, i.e. its total angular momentum (TAM) projection onto the magnetic field axis is well-defined, exactly like in the scalar QED \cite{Karlovets2023}. 
The distinction between the "evolved" photon state and the traditional "detected" state (see Refs. \cite{Bogdanov2018, Bogdanov2019, Kruining2019, Maruyama}) is similar to the difference between a polarization state that arises from the process itself and a polarization state that is measured by a detector (for instance, in Ref. \cite{BLP}). The evolved photon state is determined by a complex-valued S-matrix element $S_{fi}$, whereas the emission probability and intensity only rely on its absolute value $|S_{fi}|$. Therefore, the evolved state formalism 
clarifies the significance of the S-matrix element phase and enriches the analysis of the emission process. 

The paper is organized as follows. In Sec. \ref{Sec: Ev_state_def} the concept of the \textit{evolved states} is briefly introduced. In section \ref{Sec: Matrix_el} we derive the S-matrix element, which is further used in the analysis of the photon evolved state in Sec. \ref{Sec: Ev_state_an} and the radiation probability and intensity in Sec.\ref{Sec: Prob_int}. Finally, in Sec. \ref{Sec: Conclusion} we draw a conclusion. In Appendix \ref{Sec:El_states} we explain a straightforward approach to solving the Dirac equation in a uniform magnetic field and analyze the spin properties of the obtained electron states. 

We use natural system of units throughout the paper with $\hbar=c=1$ unless stated otherwise, whereas the electron mass and charge are denoted as $m$ and $e<0$, respectively. The Coulomb gauge is used for the four-vector potentials of the background and the radiation fields.

\section{Evolved photon state: definition}
\label{Sec: Ev_state_def}
Let us consider emission $e \rightarrow e' + \gamma$ of the photon by the electron in a constant homogeneous magnetic field. The initial state of the electron $\ket{i_e}$ and the \textit{evolved} (pre-selected) state of the final photon and electron $\ket{f}$ are connected via the first-order S-matrix in the Furry picture $\hat{S}^{(1)}$, with the interaction of the electron with the magnetic field being taken into account exactly:
\begin{equation}
    \ket{f} = \hat{S}^{(1)} \ket{i_e}.
\end{equation}
If the final electron is projected to a state $\ket{f_e}$, the state of the whole system becomes
\begin{equation}
    \ket{f'} = \sum_{\lambda=\pm 1} \int \frac{d^3 k}{(2\pi)^3}  \ket{f_e; \bm{k}, \lambda} S_{fi}^{(1)} ,
\end{equation}
where $S_{fi}^{(1)} = \bra{f_e; \bm{k}, \lambda} \hat{S}^{(1)}\ket{i_e}$ is the transition matrix element. Plane waves with the momentum $\bm{k}$ and helicity $\lambda = \pm 1$ are used here as a complete set of the one-particle photon states. As the final electron is supposed to be measured, the evolved state $\ket{f'}$ is factorized into a product of the photon state and that of the electron. Thus, the evolved state of the photon \textit{alone} is

\begin{equation}
\label{ev_state}
    \ket{\gamma}_{ev} = \sum_{\lambda=\pm 1} \int \frac{d^3 k}{(2\pi)^3}  \ket{\bm{k}, \lambda} S_{fi}^{(1)},
\end{equation}
and the corresponding photon wave function \cite{Scully} in the momentum representation becomes \cite{Karlovets2023, EPJC}

\begin{equation}
\label{A_ev}
    \bm{A}^{(ev)}(\bm{k}) = \sum_{\lambda=\pm 1} S_{fi}^{(1)} \bm{e}_{\lambda}(\bm{k}),
\end{equation}
where $\bm{e}_{\lambda}(\bm{k})$ are the polarization vectors of the plane-wave photon. The first-order S-matrix element in the Furry picture is given by the expression
\begin{equation}
\label{S_fi}
    S_{fi}^{(1)}= -ie \int d^4x\: j^\mu_{fi}(x) A_\mu^*(x),
\end{equation}
where $A_\mu$ is the four-potential of the plane-wave photon and $j^\mu_{fi}$ is the transition current. We emphasize that although the emitted photon is generally \textit{not necessarily} a plane wave, its vector potential is expressed in terms of the plane-wave transition amplitudes $S_{fi}^{(1)}$. This is simply due to the fact that we choose the plane waves as a convenient basis for expansion of the photon states. 


\section{Matrix element for photon emission}
\label{Sec: Matrix_el}
Let us now derive the S-matrix amplitude with emission of a plane-wave photon described by the state $\ket{\bm{k}, \lambda}$. 
In the Furry picture of QED, the first-order transition matrix element is given by Eq. \eqref{S_fi} where the four-potential of the plane-wave photon is
\begin{equation}
    A^\mu(x) = \frac{1}{\sqrt{2\omega V}}\,e^\mu_{\lambda} e^{-i k^\nu x_\nu}
\end{equation}
 with the polarization four-vector in the Coulomb gauge
\begin{equation}
    e^\mu_{\lambda}(\bm{k}) = \{0, \bm{e}_{\lambda}(\bm{k})\}
\end{equation}
and the wave vector
\begin{equation}
    k^\nu = \omega\{1, \sin\theta \cos\varphi_k, \sin \theta \sin \varphi_k; \cos\theta\} = \{\omega, \bm{k}\}.
\end{equation}
Note that in the cylindrical coordinates $\bm{k} = k_z\bm{e}_z + \bm{k}_{\perp}$ and $\bm{k}\cdot\bm{r} = k_z z + k_{\perp} \rho \cos(\phi_k - \phi) = k_z z + \omega r \sin\theta \cos(\phi_k - \phi)$.
The transition current reads
\begin{equation}
    j^\mu_{fi}(x) = \overline{\Psi}_f(x) \gamma^\mu \Psi_i(x),
\end{equation}
where $\Psi_i(x)$ and $\Psi_f(x)$ are the initial and the final (detected) electron states, respectively.

In the Coulomb gauge with $\bm{k} \cdot \bm{e}_{\lambda}(\bm{k})=0$, one can expand the photon polarization vectors in terms of the photon spin operator eigenfunctions as \cite{SerboUFN}
\begin{equation}
\label{e_lambda}
    \begin{aligned}
    &\bm{e}_{\lambda}(\bm{k}) = \sum_{\sigma=0, \pm1} e^{-i\sigma \varphi_k} d_{\sigma \lambda}^{(1)}(\theta)\bm{\chi}_{\sigma};\\
    &\bm{\chi}_0=(0, 0, 1),\; \bm{\chi}_{\pm 1} = \mp \frac{1}{\sqrt{2}}(1, \pm i, 0),
    \end{aligned}
\end{equation}
where $ d_{\lambda \lambda'}^{(1)}(\theta)$ are the small Wigner functions \cite{Varshalovich}.


Let us now briefly discuss the electron states that we use to calculate the transition amplitudes. We employ the stationary relativistic Landau states obtained as an exact solution to the Dirac equation in the given magnetic field in the so-called symmetric gauge with the following potential: $\bm{A} = \frac{H}{2}\{-y, x, 0\}$ (see Appendix \ref{Sec:El_states}). As the QED in the Furry picture is foremost used in problems with the very strong fields, $H \sim H_c= m^2 / |e|$, that can be found in astrophysical environments (say, in neutron stars) and the final electron stays inside \textit{the same} unperturbed field, it is natural to describe its state in terms of the same basis as the initial one. 

Although spin is a fundamental property of electron, the corresponding operator in relativistic quantum theory is defined ambiguously due to the nonuniqueness of the way how the total angular momentum operator is split into an external (orbital) part and an internal (spin) part \cite{7spin, Bliohk2017_position}. In some works it is argued that the operator obtained with the Foldy-Wouthuysen transformation is the most promising candidate for a proper relativistic spin operator \cite{7spin, Kudlis}. However, the simplest spin operator -- the so-called Pauli operator -- $\frac{1}{2}\widehat{\bm{\Sigma}}$, where
\begin{equation}
    \bm{\Sigma} = 
    \begin{pmatrix}
    \bm{\sigma} & 0  \\
     0 & \bm{\sigma} \\
  \end{pmatrix}
\end{equation}
with $\bm{\sigma}$ being the Pauli matrices, is also not devoid of meaning in relativistic theory \cite{Bliohk2017_position, 7spin}. Since the discussion of the spin-related phenomena is not the aim of this work, in the following derivations we use two orthogonal solutions, $\Psi_{s, \ell}^{\uparrow}$ and $\Psi_{s, \ell}^{\downarrow}$, which can be relatively easily obtained by the procedure described in Appendix \ref{Sec:El_states}. 
Here it is not especially significant whether these states are the eigenstates of some specific spin operator (such an operator, however, is introduced, for example, in Ref. \cite{ST}). The key point is that they posses a definite value of the \textit{total} angular momentum projection onto the field axis. In the limit of a weak magnetic field, $H \ll H_c$, they tend to the eigenstates of the spin operator $\frac{1}{2}\hat{\Sigma}_z$ with the eigenvalues $\pm\frac{1}{2}$,  that is why we refer to these states as ``spin-up'' and ``spin-down'', respectively. 

As an example, let us consider the transition between the ``spin-up'' states
\begin{equation}
\label{Psi_i}
    \Psi_i(x) = N_i^{\uparrow}
     \begin{pmatrix}
        (m+\varepsilon) \Phi_{s, \ell - 1/2}(\rho) e^{-i\varphi/2}\\
        0\\
        p_z \Phi_{s, \ell - 1/2}(\rho) e^{-i\varphi/2}\\
        -ieH \Phi_{s, \ell + 1/2}(\rho)e^{i\varphi/2}
    \end{pmatrix}
     e^{-i t \varepsilon + i \ell \varphi+i p_z z}
\end{equation}
and
\begin{equation}
\label{Psi_f}
    \Psi_f(x) = N_f^{\uparrow}
     \begin{pmatrix}
        (m+\varepsilon') \Phi_{s', \ell' - 1/2}(\rho) e^{-i\varphi/2}\\
        0\\
        p_z' \Phi_{s, \ell' - 1/2}(\rho) e^{-i\varphi/2}\\
        -ieH \Phi_{s', \ell' + 1/2}(\rho)e^{i\varphi/2}
    \end{pmatrix}
     e^{-i t \varepsilon' + i \ell' \varphi+i p_z' z},
\end{equation}
where the radial function is
\begin{equation}
    \Phi_{s, \ell}(\rho) \equiv \rho^\ell L_s^\ell(2\rho^2/\rho_H^2) e^{-\rho^2/\rho_H^2}
\end{equation}
with $\rho_H = \sqrt{\frac{4}{\abs{e}H}}$ and $L_s^{\ell}$ being the associated Laguerre polynomials.  The normalization constants $N_{i, f}$ are calculated in the Appendix \ref{Sec:El_states} (see Eq. \eqref{N_up}). The energy $\varepsilon$ is given by the equation \eqref{disp_KG}. The independent quantum numbers of this state are the continuous longitudinal momentum $p_z$, a radial quantum number $s = 0,1,2,...$ and the $z$-component of the \textit{total} angular momentum $\ell = \pm  1/2, \pm 3/2, ...$, provided $\ell \geq -s+\frac{1}{2}$. 

The corresponding transition current components (Eqs. \eqref{j0} - \eqref{j3}) and the details of the derivation are presented in the Appendix \ref{Sec: Transition currents}. After the integration we obtain the final expression for the transition amplitude \eqref{S_fi}:
\begin{multline}
\label{S_uu}
    S_{\uparrow \uparrow}^{(1)} =  ei^{-\ell+\ell'+1}(2\pi)^3\frac{N_i^{\uparrow} N_f^{\uparrow}}{\sqrt{2\omega V}}\delta(\omega + \varepsilon' - \varepsilon) \delta(k_z + p_z' - p_z) e^{i(\ell-\ell')\varphi_k} \times\\
    \times\Bigg[d_{0\lambda}^{(1)}(\theta) \rho_H^{\ell+\ell'+1}\left[m(p_z+p_z')+p_z'\varepsilon+p_z\varepsilon' \right]  F_{s, s'}^{\ell-1/2, \ell'-1/2}(y)  +\\
    4\sqrt{2}\rho_H^{\ell+\ell'}\left[(m+\varepsilon') d_{-1\lambda}^{(1)}(\theta)  F_{s, s'}^{\ell+1/2, \ell'-1/2}(y) - (m+\varepsilon) d_{1\lambda}^{(1)} (\theta)F_{s, s'}^{\ell-1/2, \ell'+1/2}(y)  \right]\Bigg].
\end{multline}
Here $y = k_{\perp}\rho_H$ and $F_{s, s'}^{\ell, \ell'}$ is defined in Eq. \eqref{eq:F}. The remaining three S-matrix elements $S_{\uparrow \downarrow}^{(1)}$, $S_{\downarrow \uparrow}^{(1)}$ and $S_{\downarrow \downarrow}^{(1)}$, which describe the transitions  $\Psi_{s, \ell}^{\uparrow} \rightarrow \Psi_{s', \ell'}^{\downarrow}$,  $\Psi_{s, \ell}^{\downarrow} \rightarrow \Psi_{s', \ell'}^{\uparrow}$ and  $\Psi_{s, \ell}^{\downarrow} \rightarrow \Psi_{s', \ell'}^{\downarrow}$ respectively, are obtained in a similar fashion. Therefore, we skip the details of the derivation and come straight to the final expressions (see Appendix \ref{Sec: Transition currents}).

\section{Evolved photon state: analysis}
\label{Sec: Ev_state_an}
Each of the four transition amplitudes obtained in the previous section can be written in a form
\begin{equation}
    \label{S_general}
    S_{fi}^{(1)} =  (2\pi)^3\delta(\omega + \varepsilon' - \varepsilon) \delta(k_z + p_z' - p_z) e^{i(\ell-\ell')\varphi_k} \mathcal{F}(\varepsilon, \varepsilon', p_z, p_z', s, s', \ell, \ell', k_{\perp}, \theta), 
\end{equation}
where $\mathcal{F}(\varepsilon, \varepsilon', p_z, p_z', s, s', \ell, \ell', k_{\perp}, \theta) = \mathcal{F}$ is some scalar function that does not depend on $\varphi_k$ (see Eqs.\eqref{S_uu}, \eqref{S_du}, \eqref{S_ud}, \eqref{S_dd}). Note that the conservation of the energy and the longitudinal momentum provided by the corresponding delta functions automatically leads to the conservation of the transverse momentum:
\begin{equation}
\label{delta_kperp}
    \delta(\omega + \varepsilon' - \varepsilon) = \delta\left(\sqrt{k_{\perp}^2 + (p_z-p_z')^2} + \varepsilon' - \varepsilon\right) = \frac{\varepsilon-\varepsilon'}{\kappa}\delta(k_{\perp}-\kappa),
\end{equation}
where
\begin{equation}
\label{kappa}
    \kappa = \sqrt{(\varepsilon-\varepsilon')^2-(p_z-p_z')^2} \geq 0
\end{equation}
is the transverse momentum of the photon. The photon spectra here coincide with those in the standard quantum synchrotron radiation theory, described, for instance, in \cite{ST, bordovitsyn1999synchrotron, Karlovets2023}.

Let us now return to the derivation of the evolved state of the emitted photon. The integral over momenta in the Eq. \eqref{ev_state} can be taken in cylindrical coordinates over $k_z$ and $k_{\perp}$ with the aid of Eq. \eqref{delta_kperp}, resulting in
\begin{equation}
\label{ev_exp}
    \ket{\gamma}_{ev} = (\varepsilon-\varepsilon') \mathcal{F} \sum_{\lambda=\pm 1} \int_0^{2\pi} d\varphi_k \ket{\bm{k}, \lambda} e^{i(\ell-\ell')\varphi_k}.
\end{equation}
In this expression, the following equalities are understood:
\begin{equation}
    \theta=\arctan\left(\frac{\kappa}{p_z-p_z'}\right),
\end{equation}
\begin{equation}
\label{k}
    \bm{k}=(\kappa \cos \varphi_k, \kappa \sin \varphi_k, p_z-p_z'),
\end{equation}
such that in the integrand of Eq. \eqref{ev_exp} only the exponential $e^{i(\ell-\ell')\varphi_k}$ and the state
$\ket{\bm{k}, \lambda}$ depend on $\varphi_k$.

Note that the expression \eqref{ev_exp} is essentially an expansion of the state $\ket{\gamma}_{ev}$ in terms of the plane wave states $\ket{\bm{k}, \lambda}$ with the same energy $\omega=\varepsilon-\varepsilon'$, longitudinal momentum $k_z = p_z-p_z'$, and a transverse momentum $\kappa$. This means that the emitted photon possesses a definite energy, definite longitudinal and transverse momenta, but not an azimuthal component of the momentum, which hints that $\ket{\gamma}_{ev}$ is a so-called \textit{Bessel beam} propagating along the field axis on average \cite{SerboUFN}. This can be confirmed by the explicit calculation of the photon TAM projection. One can easily check that due to the chosen expansion \eqref{e_lambda} of the polarization vector $\bm{e}_{\lambda}(\bm{k})$ (see discussion on this choice in Ref.\cite{EPJC}) it does not contribute to the TAM of the photon:
\begin{equation}
    \hat{j}_z^{(\gamma)}\bm{e}_{\lambda}(\bm{k}) = 0,
\end{equation}
where $\hat{\bm{j}}^{(\gamma)}=\hat{\bm{l}}^{(\gamma)}+\hat{\bm{s}}^{(\gamma)}$ is the TAM operator for a photon. Thus, the operator $\hat{j}^{(\gamma)}_z$ acts on the vector potential of the evolved state \eqref{A_ev} simply as the differentiation operator $\hat{l}^{(\gamma)}_z = -i\partial_{\varphi_k}$. Therefore, we finally find that in addition to the energy, the transverse momentum and the longitudinal one the emitted photon also has a definite TAM projection:
\begin{equation}
    \hat{j}_z^{(\gamma)}\bm{A}^{(ev)}(\bm{k}) = (\ell-\ell')\bm{A}^{(ev)}(\bm{k}).
\end{equation}
This equation illustrates that the emitted photon generally represents a twisted state rather than a plane wave. The value of $\ell-\ell'$ can, however, be zero in case of a transition with the change of the radial quantum number $s$ only. Transitions without change of both $\ell$ and $s$ are also possible, but only for the fixed value of the  longitudinal momentum $p_z'$. As we are interested in the probability integrated over $p_z'$ in the following section, such transitions are not considered here. The conservation of the TAM projection during the photon emission is of no surprise due to the azimuthal symmetry of the problem and the electron states \eqref{Psi_i} and \eqref{Psi_f} being the eigenstates of $\h{j}_z$. Importantly, the TAM of the above Bessel beam cannot be unambiguously split into the orbital part and the spin one. In this problem, this ambiguity is closely related to the same non-uniqueness of separation of the electron TAM into the OAM and the spin, which is why the operator of the latter is also ambiguously defined (see Appendix \ref{Sec:El_states}). 

We would like to stress that the Bessel beam naturally emerged in our derivations as the state of the electromagnetic field evolving directly from the process. In contrast, in Refs. \cite{Kruining2019, Maruyama, Zhang2020} the four-vector potential of a Bessel photon was explicitly substituted into the S-matrix amplitude \eqref{S_fi}, meaning that the photon was supposed to be \textit{detected} as a Bessel state with the definite parameters. An advantage of the current approach is that it allows one to derive all the parameters (the transverse momentum, the TAM, and so forth) of the Bessel beam as they evolve from the process itself instead of making an ansatz. 


The reader may well ask why we have obtained the Bessel beam, which is the simplest and somewhat idealized model of a twisted photon. Indeed, the fact that the Bessel beam naturally arises in this problem may seem surprising, because it is known to be non-integrable in the transverse plane. Since both the initial and the final states of the electron are transversely localized wave packets, one may expect the emitted photon to be localized as well, representing a normalizable Laguerre-Gaussian-like beam, or at least a weighted superposition of the Bessel states \cite{SerboUFN}. Clearly, this is not true, as well as for a scalar charged particle in magnetic field \cite{Karlovets2023}. This is because we consider the initial and final electron states possessing \textit{definite longitudinal momenta}, $p_z$ and $p_z'$ respectively, which means that they are delocalized along the z-axis. This results in a delta function $\delta(k_z+p_z'-p_z)$ in Eq. \eqref{S_general} and allows us to use Eq. \eqref{delta_kperp}, which gives rise to the definite transverse momentum of the emitted photon - a hallmark of a non-normalizable Bessel beam. 

Let us analyze what happens if the initial electron represents a longitudinally localized packet with a mean momentum value $\expv{p_z}$. For simplicity, we take a Gaussian distribution only over $p_z$ and consider the final electron with a definite $p_z'$. An example of the ``modified'' initial electron state can be
\begin{equation}
\label{Psi_i_packet}
    \displaystyle \Psi'_i(x) = \bigintss_{-\infty}^\infty dp_z \frac{1}{\sqrt{2\pi \sigma_p^2}}\, e^{-\frac{(p_z-\expv{p_z})^2}{2\sigma_p^2}} N_i
     \begin{pmatrix}
        (m+\varepsilon) \Phi_{s, \ell - 1/2}(\rho) e^{-i\varphi/2}\\
        0\\
        p_z \Phi_{s, \ell - 1/2}(\rho) e^{-i\varphi/2}\\
        -ieH \Phi_{s, \ell + 1/2}(\rho)e^{i\varphi/2}
    \end{pmatrix}
     e^{-i t \varepsilon + i \ell \varphi+i p_z z}.
\end{equation}
In this expression $\varepsilon$ depends on $p_z$ making the state nonstationary. The integration over $p_z$ in Eq. \eqref{Psi_i_packet} is successively transferred to the S-matrix element and to the evolved state. Thus, the modified evolved state simply becomes
\begin{equation}
\label{ev_packet2}
    \ket{\gamma}_{ev}' =  \frac{1}{\sqrt{2\pi \sigma_p^2}}
    \int_{-\infty}^\infty dp_z e^{-\frac{(p_z-\expv{p_z})^2}{2\sigma_p^2}} \ket{\gamma}_{ev}.
\end{equation}
Note that here $\ket{\gamma}_{ev}$ given by the Eq. \eqref{ev_exp} is a function of $p_z$. Since $\kappa$ included in the expression for $\ket{\gamma}_{ev}$ depends on $p_z$, $\ket{\gamma}_{ev}'$ represents an intricate superposition of the Bessel beams with some distribution over both the longitudinal and transverse momenta. Although it is impossible to explicitly factor out the Gaussian distribution over the $\kappa(p_z)$ in this expression, it is tempting to say that Eq.\eqref{ev_packet2} is a Bessel-Gauss-like state, which is normalizable \cite{SerboUFN}. Thus, the ``unphysical'' Bessel beam would not emerge in our derivation if we considered the nonstationary and longitudinally localized states of the electron. This underlines the importance of nonstationary states for the correct description of the electron in an external field, which has already been noted in \cite{Sizykh} and applied to the dynamics of the electron r.m.s. radius in a solenoid. 

\section{Emission probability and intensity}
\label{Sec: Prob_int}
Having derived and analyzed the state $\ket{\gamma}_{ev}$ in which the photon is emitted, we now turn to the radiation probability. Recalling the expression \eqref{ev_state} for the evolved state of the photon we can find the probability of transition between the electron states with some fixed quantum numbers as

\begin{align}
\label{dW_sl(p_z)}
    &W^{(1)}_{s', \ell'}(p_z') = \leftindex_{ev}{\braket{\gamma}}_{ev} = \sum_{\lambda, \lambda'=\pm1} \int \frac{d^3 k}{(2\pi)^3} \frac{d^3 k'}{(2\pi)^3} \bra{\bm{k}',\lambda'}\ket{\bm{k}, \lambda} S_{fi}^{(1)}(\bm{k}, \lambda) S_{fi}^{(1)*}(\bm{k}', \lambda') = \\
    &=\sum_{\lambda=\pm1} \int \frac{d^3 k}{(2\pi)^3} \left|S_{fi}^{(1)}\right|^2,
\end{align}
where we have used the orthogonality relation for plane waves.

The total radiation probability is then given by the summation over all possible quantum numbers of the final electron state ($s', \ell'$) and the integration over the final longitudinal momentum $p_z'$:
\begin{equation}
    W^{(1)} = \sum_{s', \ell'}\int  W^{(1)}_{s', \ell'}(p_z')\frac{dp_z'}{2\pi}L,
\end{equation}
where $L$ is the normalization length. However, since we are interested in the emission of the photons with definite values of the TAM, we will not sum over the discrete quantum numbers and we only integrate Eq. \eqref{dW_sl(p_z)} over $p_z'$ introducing the following probability:
\begin{equation}
    W_{s', \ell'}^{(1)} \equiv  \int L\frac{dp_z'}{2\pi}  \; W^{(1)}_{s', \ell'}(p_z')  = \sum_{\lambda=\pm1} \int \frac{dp_z'}{2\pi} L \frac{d^3 k}{(2\pi)^3} \left|S_{fi}^{(1)}\right|^2.
\end{equation}

When squaring the delta functions in the matrix element, we use the rule

\begin{equation}
    (\delta(\omega+\varepsilon'-\varepsilon))^2(\delta(p_z-p_z'-k_z))^2 \rightarrow \frac{T}{2\pi}\delta(\omega+\varepsilon'-\varepsilon)\frac{L}{2\pi}\delta(p_z-p_z'-k_z),
\end{equation}
and for the integration over $d^3 k$ in cylindrical coordinates we use Eq. \eqref{delta_kperp}. The corresponding  emission
probability per unit time is found as
\begin{equation}
\label{dW(p_z)}
     \dot{W}^{(1)}_{s', \ell'} = \frac{W^{(1)}_{s', \ell'}}{T} =  \frac{1}{(2\pi)^4} L\int dp_z'\int k_{\perp} d k_{\perp} d\varphi_k dk_z \sum_{\lambda=\pm 1}\left|S_{fi}^{(1)}\right|^2.    
\end{equation}
For instance, for the ``up-up'' (no spin-flip) transition substituting the amplitude \eqref{S_uu} into the Eq. \eqref{dW(p_z)} yields
\begin{multline}
\label{W_uu}
     \dot{W}^{(1)}_{s', \ell'} = \pi L^2 e^2 N_i^2\int d p_z' N_f^2 \times\\
     \sum_{\lambda=\pm 1}\Bigg|d_{0\lambda}^{(1)}(\theta) \rho_H^{\ell+\ell'+1}\left[m(p_z+p_z')+p_z'\varepsilon+p_z\varepsilon' \right]  F_{s, s'}^{\ell-1/2, \ell'-1/2}(y)  +\\
    4\sqrt{2}\rho_H^{\ell+\ell'}\left[(m+\varepsilon') d_{-1\lambda}^{(1)}(\theta)  F_{s, s'}^{\ell+1/2, \ell'-1/2}(y) - (m+\varepsilon) d_{1\lambda}^{(1)} (\theta)F_{s, s'}^{\ell-1/2, \ell'+1/2}(y)  \right]\Bigg|^2,
\end{multline}
where $N_f$, $\varepsilon'$, $\sin\theta = \kappa/\omega$, and $\kappa$ depend on $p_z'$.

\begin{figure}[h]
    \centering
    \includegraphics[width=\textwidth]{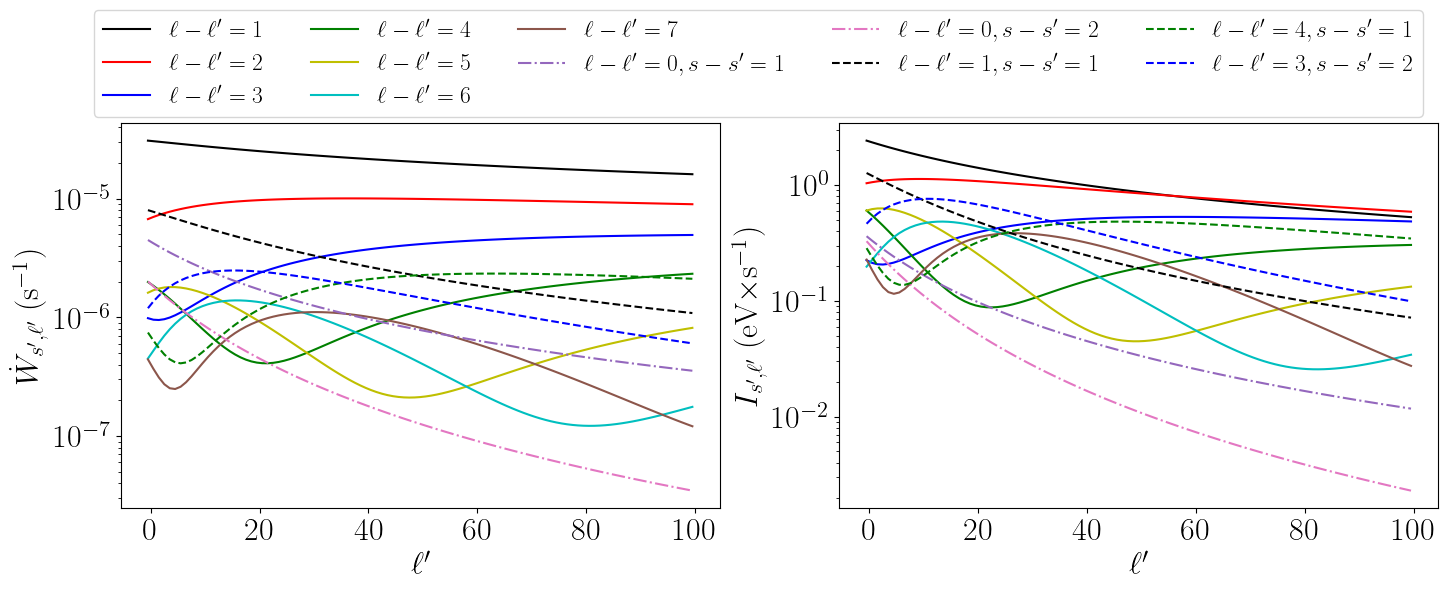} 
     \caption{The emission probability \eqref{W_uu} (left) and the corresponding intensity \eqref{Intensity} (right) for $H=H_c$, $p_z=10^{-3} mc$, and no spin-flip. For the solid lines $s = s' = 20$, the dashed lines correspond to the twisted photons with a simultaneous change of the radial quantum number $s \to s'\ne s$, the dash dotted lines correspond to the untwisted photons with the TAM $j_z=\ell-\ell'=0$.}
     \label{plot_1}
\end{figure}

\begin{figure}[h]
    \centering
    \includegraphics[width=\textwidth]{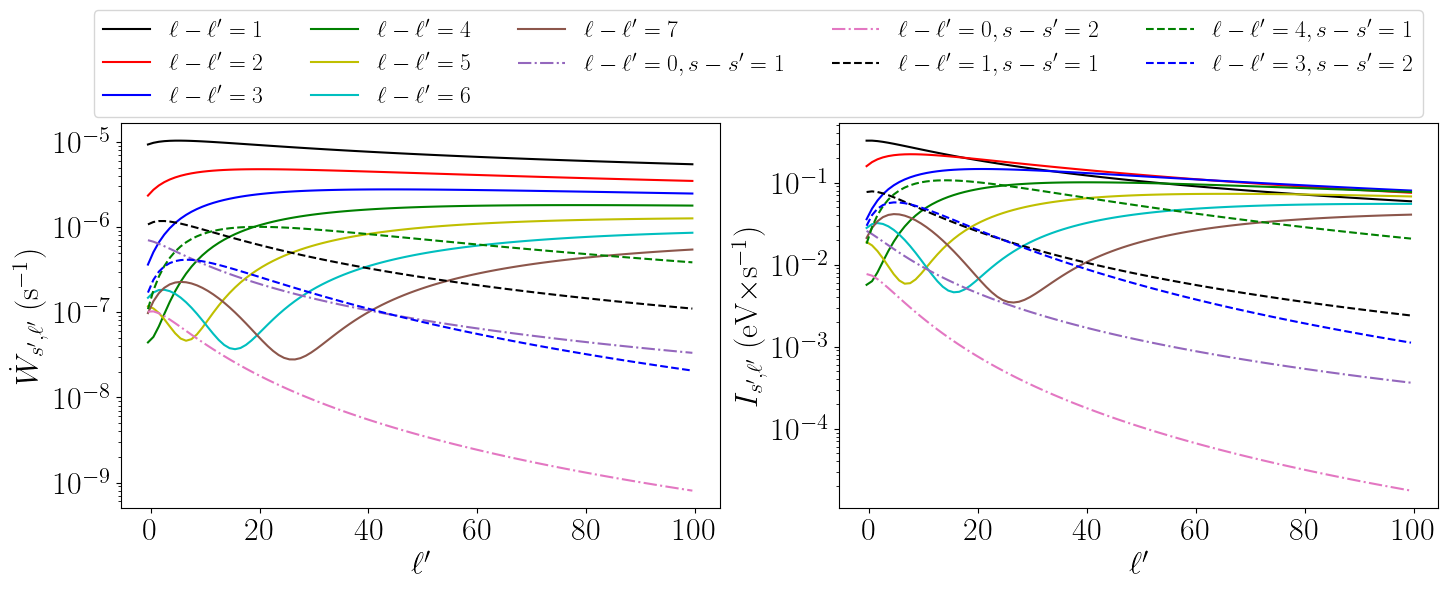}
     \caption{The same as in Fig \ref{plot_1}, but for $H=0.1H_c$ and $s=5$.}
     \label{plot_2}
\end{figure}

\begin{figure}[h]
    \centering
    \includegraphics[width=\textwidth]{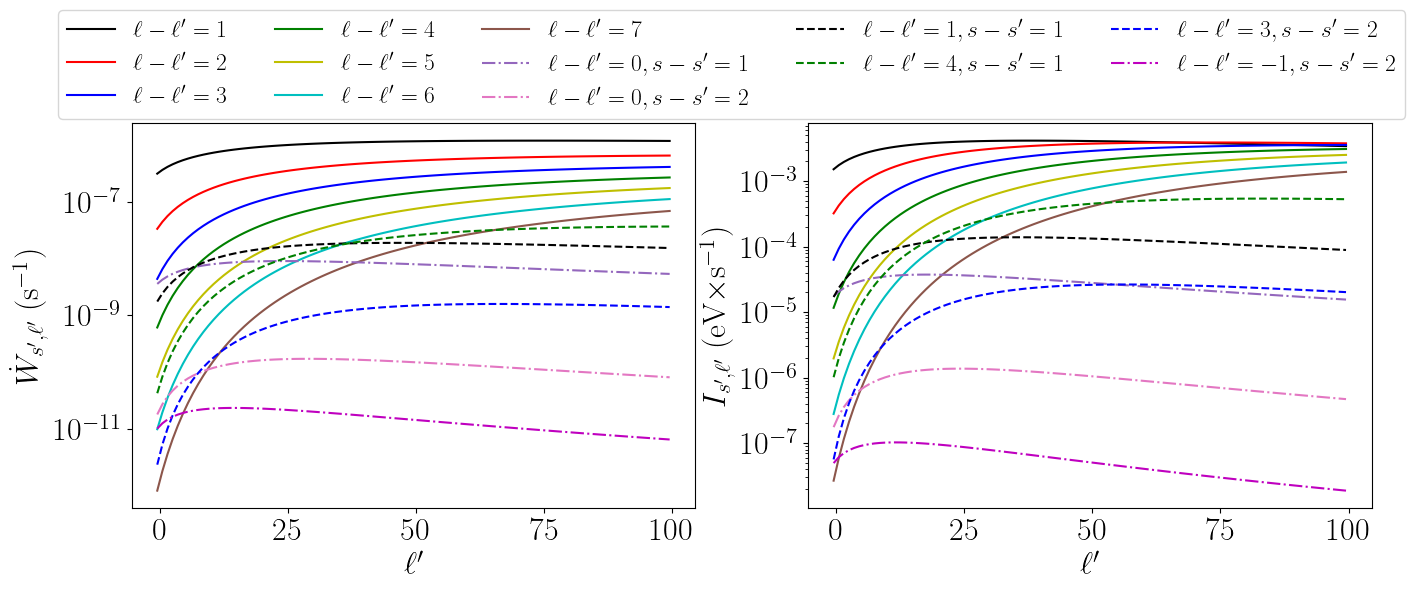}
     \caption{
     The emission probability \eqref{W_uu} (left) and the corresponding intensity \eqref{Intensity} (right) for $H=0.01H_c$, $p_z=10^{-3} mc$, and no spin-flip. For the solid lines $s = s' = 5$, the dashed lines correspond to the twisted photons with a simultaneous change of the radial quantum number $s \to s'\ne s$. The magenta dash-dotted line corresponds to the transitions with increase of the electron OAM, so that the photon TAM is $\ell-\ell'=-1$. Other dash dotted lines correspond to the untwisted photons with $j_z=\ell-\ell'=0$.}
     \label{plot_3}
\end{figure}

Correspondingly, the radiation intensity is obtained by multiplying the integrand by the photon energy $\omega = \varepsilon-\varepsilon'$, which also depends on $p_z'$:
\begin{equation}
\label{Intensity}
   I_{s', \ell'}^{(1)} \equiv  \int L \frac{dp_z'}{2\pi}  \; \omega \dot{W}^{(1)}_{s', \ell'}(p_z').
\end{equation}
The integration over $p_z'$ in equations \eqref{W_uu} and \eqref{Intensity}  is to be carried out numerically over the region allowed by the condition $\kappa \geq0$ (see Eq. \eqref{kappa} and Ref.\cite{Karlovets2023} for more detail):
\begin{equation}
    p_z' \in \left[p_z-\frac{p_{\perp}^2-(p_{\perp}')^2}{2(\varepsilon-p_z)}, p_z+\frac{p_{\perp}^2-(p_{\perp}')^2}{2(\varepsilon+p_z)} \right].
\end{equation}
The radiation probability $\dot{W}^{(1)}_{s', \ell'}$ for three other types of possible transitions can be obtained similarly from the amplitudes \eqref{S_ud} - \eqref{S_dd}. For the sake of brevity we do not present the corresponding expressions here.

\begin{figure}[h]
    \centering
    \includegraphics[width=\textwidth]{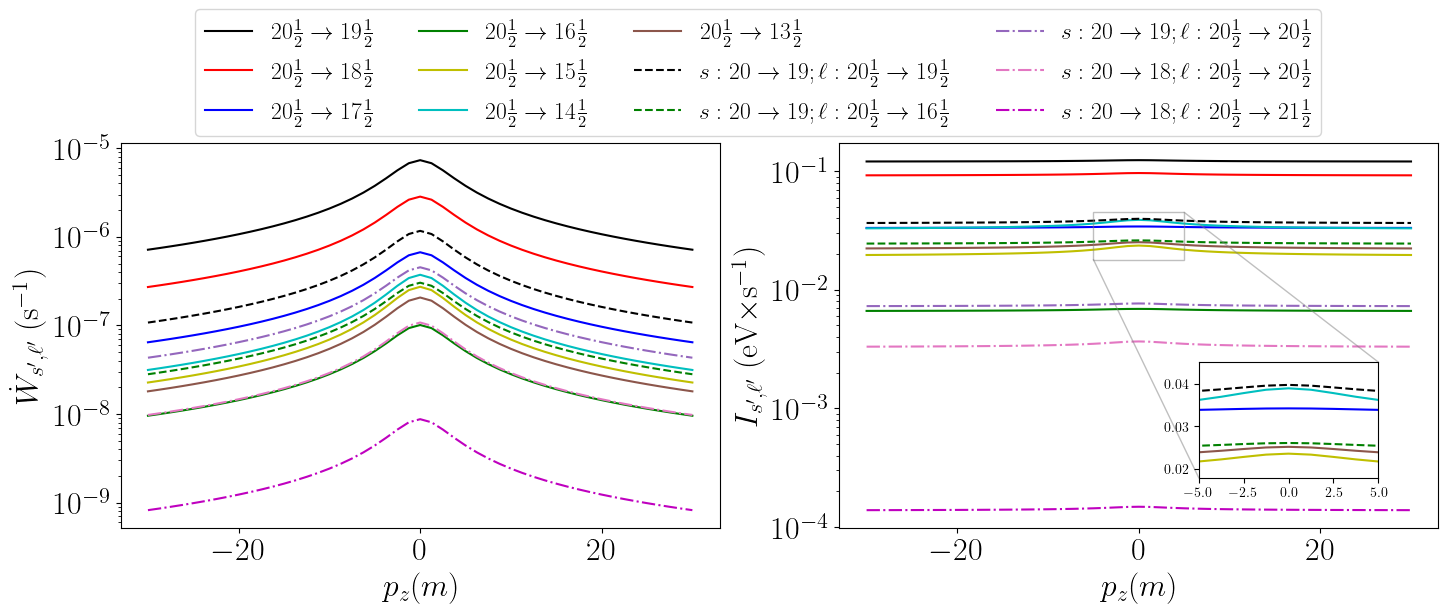}
     \caption{The dependence of the emission probability (left) and the intensity (right) on the electron momentum $p_z$ for $H=0.1H_c$, $s=s'=20$. The transition $20\frac{1}{2} \to 19\frac{1}{2}$ means $\ell = 20\frac{1}{2}$, $\ell' = 19\frac{1}{2}$, $s = s' = 20$; those with $\ell : 20\frac{1}{2} \to 20\frac{1}{2}$ correspond to the untwisted photons with $j_z=0$. The green line overlaps with the pink dashed one on the left; the cyan line on the left overlaps with the blue one on the right. The magenta dash-dotted line corresponds to increase of the electron OAM during the emission (so that the photon TAM is $\ell-\ell'=-1$).}
     \label{plot_5}
\end{figure}

Due to dependence of the radiated photon energy on $\ell$ and $\ell'$, the emission probability and the intensity depend slightly differently on $\ell'$. For ``up-up'' transitions this dependence is shown in Fig. \ref{plot_1}, Fig. \ref{plot_2} and Fig. \ref{plot_3}. Electron is supposed to be nonrelativistic with $p_z = 10^{-3}m$ and $H = H_c$, $H = 10^{-1} H_c$, and $H = 10^{-2} H_c$, respectively. Note that the results in Figs. \ref{plot_1}, \ref{plot_2} and \ref{plot_5} are presented for the same parameters as in Ref. \cite{Karlovets2023} for a spinless particle. It is important to note that the absolute values of probability and intensity agree with the predictions of the scalar QED within at least 15\% while qualitatively the dependencies are absolutely the same. While the probability of the transitions $\ell \to \ell'=\ell-1$ always dominates, their intensity becomes slightly less than that of $\ell \to \ell'=\ell-2$ starting from $\ell' \geq 20$ at $H \sim 0.1 H_c$. Moreover, in subcritical magnetic field $H=0.1 H_c$ even the intensities of the transitions $\ell \to \ell'=\ell-3, \ell-4$ also exceed that for $\ell \to \ell'=\ell-1$ starting from $\ell' \geq 50$ (see Fig. \ref{plot_2}). As can be seen from the Figs. \ref{plot_1}-\ref{plot_3}, the transitions with radiation of ``untwisted'' photons are generally suppressed compared to those with $\ell \to \ell' \ne \ell$. The transitions with the increase of the electron TAM $\ell\to \ell' > \ell$ are even more damped, as demonstrated in Fig. \ref{plot_3}.

\begin{figure}[h]
	\centering
	\begin{subfigure}{0.45\linewidth}
		\includegraphics[width=\linewidth]{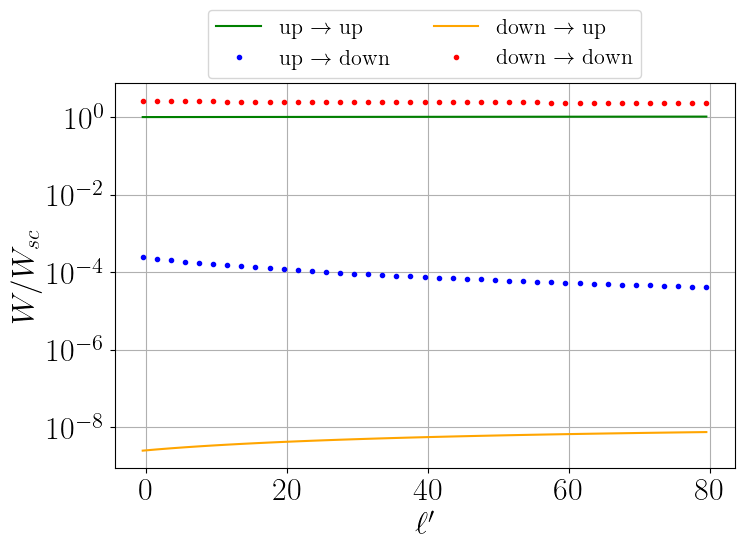}
	\end{subfigure}
	\begin{subfigure}{0.45\linewidth}
		\includegraphics[width=\linewidth]{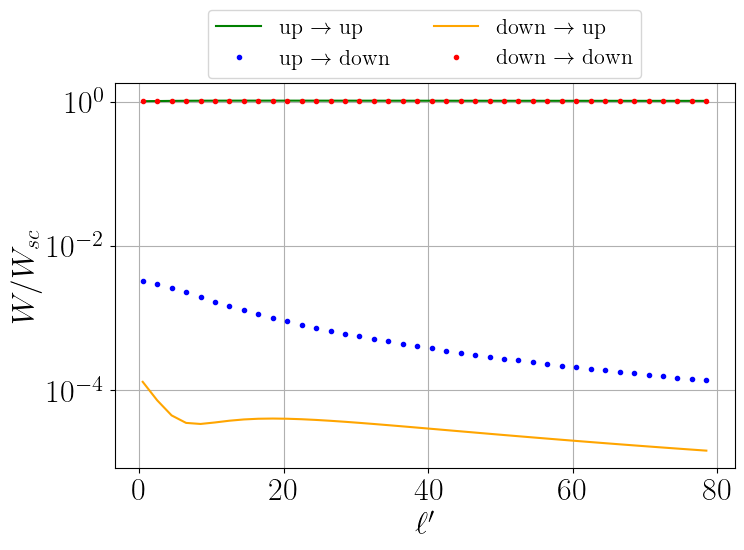}
	\end{subfigure}
 
	\caption{The ratio of four possible types of transition probabilities to the probability of emission by a scalar charge derived in \cite{Karlovets2023}. $H=10^{-3}H_c$ (left) and $H=H_c$ (right); $\ell-\ell'=3$, $s=s'=20$ for all transitions.}
	\label{ratios}
\end{figure}

Let us now evaluate the probabilities of three other types  of the transitions -- ``up-down'', ``down-up'' and ``down-down'' -- and compare these results with those obtained in \cite{Karlovets2023}, where the same process is considered for a scalar charged particle.  Since the initial and final states of a spinless particle in magnetic field are characterized by the \textit{integer} values of TAM ($\ell$ and $\ell'$, respectively), it seems sensible to compare a ``scalar'' transition $\ell \to \ell'$ with the ``spin'' transitions  $\ell+1/2 \to \ell'+1/2$ (shown in Fig. \ref{ratios}) or $\ell-1/2 \to \ell'-1/2$. The latter comparison is not presented in the figures as it does not qualitatively differ from the former. Thus, we consider the ratio $\dot{W}^{(1)}_{s', \ell'} / \dot{\mathcal{W}}^{(1)}_{s', \ell' + 1/2}$, where $\dot{\mathcal{W}}^{(1)}_{s', \ell'}$ is the radiation probability per unit time for a scalar charge. Two regimes are presented in figure \ref{ratios}: $H=10^{-3}H_c$ (left) and $H=H_c$ (right). It can be seen that generally the probabilities of ``up-up'' and ``down-down'' transitions are both of the same order and are in good agreement with the predictions of the scalar QED. On the contrary, the probabilities of ``up-down'' and ``down-up'' transitions are significantly suppressed as compared to the transitions without the spin flip. The ``up-down'' probabilities, in turn, exceed those for the inverse spin flip by many orders, which is a manifestation of a well-know effect of self-polarization predicted by Sokolov and Ternov \cite{ST} and lately investigated for the case of weakly excited electron states in ultrastrong magnetic fields \cite{Kruining2019}. Due to the spin-flip transitions being considerably damped, the probabilities of ``up-up'' transitions shown in Figs. \ref{plot_1} - \ref{plot_3} can be viewed as summed over the spin states of the final electron with a good accuracy.

Fig. \ref{plot_5} illustrates the radiation probability per second and the intensity for the definite discrete quantum numbers of the final electron as a function of the initial momentum $p_z$ along the magnetic field. Similarly to Figs. \ref{plot_1} - \ref{plot_3} we consider here only ``up-up'' transitions. Both probability and intensity turn out to be symmetric functions of $p_z$ with a maximum at $p_z=0$ although the bispinors \eqref{Psi_Plus} and \eqref{Psi_Minus} depend on the sign of $p_z$. Whereas the momentum-dependent probabilities for the untwisted photons emission are not necessarily the lowest ones, the intensities are almost independent on the momentum $p_z$ and demonstrate the domination of the twisted radiation with the TAM $\ell-\ell' \sim 1, 2$ and unchanged $s$. In scalar QED, however, the radiation intensity was shown to be completely independent on $p_z$ \cite{Karlovets2023}.

\section{Discussion and conclusion}
\label{Sec: Conclusion}
The dynamics of electrons in magnetic fields and the synchrotron radiation have been studied for a very long time both in classical and quantum formalisms. However, the transfer of the quantized angular momentum from electrons to the electromagnetic field has only recently been discovered and analyzed within classical \cite{katoh2017helical, katoh2017angular, epp2019angular, epp2022angular} and quantum \cite{Epp2023, Karlovets2023} frameworks. Here, we have ascertained that the definite TAM projection is a natural property of the photon state evolving from the process itself rather than a consequence of the choice of the detector.

We have found that the vast majority of the emitted photons are indeed twisted, representing the Bessel beams with the TAM projection $j_z = \ell-\ell'$, although a small part of them are emitted due to the transitions without a change of the angular momentum. The radiation of such non-twisted photons 
is not predicted by classical electrodynamics. 
For critical fields, $H \lesssim H_c$, the energies of the twisted photons can belong to the hard X-ray and $\gamma$-range \cite{Karlovets2023}. 

We have also argued that the reason why the transversely localized electron packet emits a delocalized Bessel photon is that here we do not account for the longitudinal localization of the electron. Considering the longitudinally localized packets would lead to the radiation of fully localized, yet diffractively spreading photon beams. Spatial localization of the electron Landau states in all three dimensions can be especially important for proper description of the transitions in metals and solids exposed to magnetic fields, as well as in Penning traps, because the resultant twisted photon packets will also be localized. Say, in the critical magnetic fields their transverse coherence length can be as small as of the order of the electron Compton wavelength \cite{Karlovets2023}. Depending on the spatial coherence, these twisted photons can interact differently with other electrons, ions, phonons, cosmic plasma, and so forth.

Although the spin states of the electron in magnetic field can be defined ambiguously, the emitted photons turn out to be twisted regardless of the electron polarization description. The choice of the concrete spin operator corresponds to the specific separation of the TAM operator, the latter being uniquely defined, into its orbital and spin parts. However, it is the TAM projection of the photon that is observable in this problem and its value is determined only by the TAM of the initial and final electron states, which \textit{do not} depend on the choice of the electron spin operator. The differential probability and intensity of transitions, nevertheless, depend on the electron polarization revealing the self-polarization effect. 

In conclusion, we note that the angular momentum of the emitted photons does not usually manifest itself at the terrestrial experiments (e.g. at storage rings) as it is quantized along the field, while the photons are being detected at the angles close to the orbital plane where the wave front looks almost flat. The helical structure of the wave front can be noticed only when observing the radiation at angles close to the magnetic field axis, especially in the critical and subcritical fields, $H \lesssim H_c$, typical for magnetars.
The importance of twisted photons in astrophysics lies primarily in their ability to interact with the medium  differently from the plane waves, e.g., to excite the Landau electrons to higher angular momentum states or induce stimulated emission, which can have implications for studying stellar nucleosynthesis \cite{Maruyama}. Finally, the dynamics of the spreading photon packets emitted by the localized electron states -- say, during the propagation through the interstellar medium -- may also be an interesting problem because the cross section of elastic scattering of twisted photons on charged particles usually has a maximum at \textit{much larger angles} than it is for the plane-wave photons. 


\

We are thankful to S.\,Baturin, A.\,Chaikovskaia, D.\,Glazov, D.\,Grosman, G.\,Sizykh, A.\,Volotka, and, especially, to A.\ Di Piazza for useful discussions and criticism.
The studies in Sec. II are supported by the Government of the Russian Federation through the ITMO Fellowship and Professorship Program.
The studies in Sec. III are supported
by the Russian Science Foundation (Project No. 21-42-04412; https://rscf.ru/en/project/21-42-04412/). 
The work on the evolved state in Sec.IV (by D. Karlovets) was supported by the Foundation for the Advancement of Theoretical Physics and Mathematics “BASIS”. The studies in Sec.V are supported by the Russian Science Foundation (Project No.\,23-62-10026;  https://rscf.ru/en/project/23-62-10026/).

\


\appendix

\section{Appendix: Electron states in magnetic field}
\label{Sec:El_states}
\subsection{Solution of the Dirac equation}
In this section, we describe a simple method of obtaining a solution of Dirac equation in a constant homogeneous magnetic field 
utilizing an approach of \cite{BLP}. Slightly different realization of the procedure described below was lately proposed in \cite{Kruining2017}.

First, let us define the Klein-Gordon operator:
\begin{equation}
    \h{K} \equiv (\hp-e\mathcal{A})^2 - m^2 = (i\partial^\mu - e\mathcal{A}^\mu)^2 - m^2 = -\square - 2ie \mathcal{A}_\mu \partial^\mu + e^2\mathcal{A}^2 - m^2.
\end{equation}
The Klein-Gordon equation then simply reads as

\begin{equation}
    \h{K} \Phi(x) = 0.
\end{equation}
We consider constant  magnetic field $\bm{H} = \{0, 0, H\}$, which can be described by a vector-potential in two different gauges -- the Landau gauge and the symmetric one -- resulting in the Hermite-Gaussian beams and in the axially symmetric Laguerre-Gaussian packets, respectively. We choose the four-vector potential in the later symmetric gauge:
\begin{equation}
\begin{aligned}
    &\mathcal{A}^{\mu} 
    = \{0, \bm{A}\};\\
    &\bm{A} = \frac{H\rho}{2}\bm{e}_{\varphi} = \frac{H}{2}\{-y, x, 0\}.
\end{aligned}
\end{equation}
Then the Klein-Gordon in cylindrical coordinates takes the following form:
\begin{equation}
     \left[-\partial_t^2 + \frac{1}{\rho}\frac{\partial}{\partial \rho}\left(\rho \frac{\partial}{\partial \rho}\right) + \frac{1}{\rho^2}\frac{\partial^2}{\partial\varphi^2} + \partial_z^2 - ieH\partial_{\varphi} - e^2\frac{H^2\rho^2}{4}-m^2\right]\Phi(x) = 0.
\end{equation}
By taking the following ansatz for the solution 
\begin{equation}
    \Phi(x) \equiv \Phi(\rho) e^{-i \varepsilon t + i \ell \varphi + ip_z z}
\end{equation}
we obtain the equation for the transverse part of the wave function:
\begin{equation}
   \left[\partial_{\rho}^2 + \frac{1}{\rho}\partial_{\rho} - \frac{\ell^2}{\rho^2} + e \ell H - \left(\frac{eH\rho}{2}\right)^2+\varepsilon^2-p_z^2-m^2\right]\Phi(\rho) = 0.
\end{equation}
Following \cite{ST}, the solution of (6) can be written as

\begin{equation}
    \Phi(\rho) = \mathcal{N}\left(\frac{2\rho^2}{\rho_H^2}\right)^{\ell/2} L_s^\ell(2\rho^2/\rho_H^2)  e^{-\rho^2/\rho_H^2},
\end{equation}
where $\mathcal{N}$ is a normalization constant, $\rho_H = \sqrt{\frac{4}{\abs{e}H}}$, $L_s^{\ell}$ are the associated Laguerre polynomials and the energy dispersion is
\begin{equation}
\label{disp_KG}
    \varepsilon^2_{s, \ell} - m^2 - p_z^2 \equiv p_{\perp}^2 = 2\abs{e}H(s+\ell+1/2).
\end{equation}
Let us define the transverse function without the prefactor as
\begin{equation}
    \Phi_{s, \ell}(\rho) \equiv \rho^\ell L_s^\ell(2\rho^2/\rho_H^2) e^{-\rho^2/\rho_H^2}.
\end{equation}
Second, we consider the Dirac equation
\begin{equation}
\label{Eq:Dirac}
    \left[\gamma^{\mu} (\hp_{\mu} - eA_{\mu})-m\right]\Psi(x) = 0,
\end{equation}
where $\gamma^{\mu}$ are the gamma matrices in the standard representation. Instead of solving the first-order equation \eqref{Eq:Dirac} directly, one can first transform it into a second-order equation by applying the ``projection'' operator $\gamma^{\mu} (\hp_{\mu} - e A_{\mu})+m$ \cite{BLP}:
\begin{equation}
\label{Eq:Dirac_expl}
    \left[(\hp - eA)^2 - m^2 + e\bm{\Sigma}\cdot \bm{H} - i e \bm{\alpha} \cdot \bm{E}\right]\Phi(x) = 0.
\end{equation}
Here matrices $\bm{\alpha}$ and $\bm{\Sigma}$ are defined as follows:
\begin{equation}
    \bm{\alpha} = 
    \begin{pmatrix}
    0 & \bm{\sigma} \\
     \bm{\sigma} & 0 \\
  \end{pmatrix}, \; \bm{\Sigma} = 
    \begin{pmatrix}
    \bm{\sigma} & 0  \\
     0 & \bm{\sigma} \\
  \end{pmatrix},
\end{equation}
where $\bm{\sigma}$ are the Pauli matrices. The third and the fourth terms here arise from the products of gamma-matrices and the electromagnetic tensor. The solution of the first-order equation $\Psi(x)$ is then obtained by applying the projection operator to $\Phi(x)$:
\begin{equation}
    \Psi(x) = \left[\gamma^{\mu} (\hp_{\mu} - e A_{\mu})+m\right]\Phi(x)
\end{equation}
Substituting the expression for the magnetic field to \eqref{Eq:Dirac_expl} we get
\begin{equation}
    \left[(\hp - eA)^2 - m^2 + e H 
    \begin{pmatrix}
    1 & 0 & 0& 0\\
     0 & -1 & 0& 0\\
     0 & 0 & 1& 0\\
     0 & 0 & 0& -1
  \end{pmatrix}\right]\Phi(x) = 0.
\end{equation}
This is a system of four Klein-Gordon-like equations
\begin{equation}
\label{Eq:4KG}
    \left[\h{K}\begin{pmatrix}
    1 & 0 & 0& 0\\
     0 & 1 & 0& 0\\
     0 & 0 & 1& 0\\
     0 & 0 & 0& 1
  \end{pmatrix} + e H 
    \begin{pmatrix}
    1 & 0 & 0& 0\\
     0 & -1 & 0& 0\\
     0 & 0 & 1& 0\\
     0 & 0 & 0& -1
  \end{pmatrix}\right]\Phi(x) = 0,
\end{equation}
that leads to two independent equations
\begin{equation}
    \begin{cases}
        (\h{K} + e H)\Phi_{+}^{KG} = 0\\
        (\h{K} - e H)\Phi_{-}^{KG} = 0.
    \end{cases}
\end{equation}
Then the solution of Eq. \eqref{Eq:4KG} is
\begin{equation}
\label{sol_with_c_i}
    \Phi(x) \equiv 
    \begin{pmatrix}
        c_1 \Phi_{+}^{KG}\\
        c_2 \Phi_{-}^{KG}\\
        c_3 \Phi_{+}^{KG}\\
        c_4 \Phi_{-}^{KG}
    \end{pmatrix}.
\end{equation}
where $c_i$ are the arbitrary complex-valued constants.
A naive choice of $\Phi_{\pm}^{KG}$ could be
\begin{equation}
    \Phi_{\pm}^{KG} =  \Phi_{s, l}(\rho)e^{-i t \varepsilon_{s, l\mp 1/2} + i l \varphi+i p_z z},
\end{equation}
However, such a solution is not physically relevant because $\Phi_{+}^{KG}$ and $\Phi_{-}^{KG}$ should correspond to the same energy. To satisfy this condition, one should choose $\Phi_{\pm}^{KG}$ in another form:
\begin{equation}
    \Phi_{\pm}^{KG} =  \Phi_{s, \ell\mp 1/2}(\rho)e^{-i t \varepsilon_{s, \ell} + i (\ell\mp1/2) \varphi+i p_z z}
\end{equation}
(here $\ell$ is a \textit{half-integer}).

First, we consider the following two ``auxillary'' solutions of \eqref{Eq:4KG}:
\begin{equation}
\label{Phi_Plus}
     \Phi_{s, \ell}^{\uparrow}(x) = N^{\uparrow}\Phi_{s, \ell - 1/2}(\rho)e^{-i t \varepsilon_{s, \ell} + i (\ell-1/2) \varphi+i p_z z} \begin{pmatrix}
        1\\
        0\\
        0\\
        0
    \end{pmatrix}
\end{equation}
and 
\begin{equation}
\label{Phi_Minus}
     \Phi_{s, \ell}^{\downarrow}(x) = N^{\downarrow}\Phi_{s, \ell + 1/2}(\rho)e^{-i t \varepsilon_{s, \ell} + i (\ell+1/2) \varphi+i p_z z} \begin{pmatrix}
        0\\
        1\\
        0\\
        0
    \end{pmatrix}
\end{equation}
(the notation $\uparrow$ and $\downarrow$ will become clear later on). These are clearly the eigenfunctions of the operators $\frac{1}{2}\widehat{\Sigma}_z$, $\hat{l}_z = -i\partial_\varphi$ and $\hat{j}_z = \frac{1}{2}\widehat{\Sigma}_z +\hat{l}_z $, but not yet the solutions of the Dirac equation \eqref{Eq:Dirac}. By acting with the projection operator on Eq. \eqref{Phi_Plus} and using the relations
\begin{equation}
    \begin{aligned}
        & x L_s^\ell(x) = (s+\ell)L_s^{\ell-1}(x) - (s+1) L_{s+1}^{\ell-1}(x); \\
        & L_s^\ell(x) = L_s^{\ell+1}(x)-L_{s-1}^{\ell+1}(x);\\
        & \frac{d}{dx} L_s^\ell(x) = -L_{s-1}^{\ell+1}(x),
    \end{aligned}
\end{equation}
we obtain the following solution of the Dirac equation:
\begin{equation}
\begin{aligned}
\label{Psi_Plus}
   & \Psi_{s, \ell}^{\uparrow}(x) = N^{\uparrow}
    \begin{pmatrix}
        (m+\varepsilon) \Phi_{s, \ell - 1/2}(\rho) e^{-i\varphi/2}\\
        0\\
        p_z \Phi_{s, \ell - 1/2}(\rho) e^{-i\varphi/2}\\
        -ieH \Phi_{s, \ell + 1/2}(\rho)e^{i\varphi/2}
    \end{pmatrix}
     e^{-i t \varepsilon_{s, \ell} + i \ell \varphi+i p_z z};\\
     &\varepsilon_{s, l} = \sqrt{m^2+p_z^2+p_{\perp}^2},\: p_{\perp}^2 = \frac{8}{\rho_H^2}(s+\ell+1/2) = 2 m^2 \frac{H}{H_c} (s+\ell+1/2).
\end{aligned}
\end{equation}
One can similarly consider second auxiliary solution: after applying the projector to Eq. \eqref{Phi_Minus} we find
\begin{equation} 
\label{Psi_Minus}
    \Psi_{s, \ell}^{\downarrow}(x) = N^{\downarrow}
    \begin{pmatrix}
        0\\
         (m+\varepsilon) \Phi_{s, \ell + 1/2}(\rho)e^{i\varphi/2}\\
        -2i(\ell+s+1/2)\Phi_{s, \ell - 1/2}(\rho) e^{-i\varphi/2}\\
        -p_z \Phi_{s, \ell + 1/2}(\rho)e^{i\varphi/2}
    \end{pmatrix}
     e^{-i t \varepsilon_{s, \ell} + i \ell \varphi+i p_z z}.
\end{equation}
Notably, the energy $\varepsilon_{s, \ell}$ here is the same as that in Eq. \eqref{Psi_Plus}: it is determined \textit{by the TAM projection} $\ell$ rather than by the spin or the orbital momentum separately. These states are exactly the solutions obtained in \cite{ST} with another method and 
coincide with those described in \cite{Kruining2017} when the TAM projection $\ell$ is positive. We employ these states for all the calculations in the main text of the paper. Let us normalize the states as follows:
\begin{equation}
\int d^3x \: j^0(x) = \int_0^{2\pi} d\varphi \int_{-L/2}^{L/2} dz \int_0^{\infty} \rho d\rho \Psi_{s, \ell}^{\uparrow}(x)^\dag \Psi_{s, \ell}^{\uparrow}(x) = 1.
\end{equation}
Note that according to Eq. (7.414) in \cite{Gr-R}
\begin{equation}
    \int_0^\infty \Phi_{s, \ell}^2(\rho) \rho d\rho = \frac{\rho_H^{2\ell+2}}{2^{\ell+2}}\frac{(s+\ell)!}{s!}.
\end{equation}
Then the normalization constants are
\begin{equation}
\label{N_up}
    N^{\uparrow} = \sqrt{\frac{2^{\ell-1/2}s!}{L \pi(s+\ell-1/2)!}}\frac{\rho_H^{-\ell-1/2}}{\sqrt{\varepsilon(\varepsilon+m)}},
\end{equation}
\begin{equation}
    N^{\downarrow} = \sqrt{\frac{2^{\ell+1/2}s!}{L \pi(s+\ell+1/2)!}}\frac{\rho_H^{-\ell-3/2}}{\sqrt{\varepsilon(\varepsilon+m)}}.
\end{equation}
One can easily check that the states $\Psi_{s, \ell}^{\uparrow}$ and $\Psi_{s', \ell'}^{\downarrow}$ are orthogonal in the sense of the integral
\begin{equation}
    \label{orth_0}
    \int_0^\infty \rho d\rho \Psi_{s, \ell}^{\uparrow}(x)^\dag \Psi_{s', \ell'}^{\downarrow}(x) = 0
\end{equation}
even for $s=s'$ and $\ell = \ell'$. The states of the same type with the different quantum numbers $s$, $\ell$ and $p_z$ are also orthogonal:
\begin{equation}
    \label{orth}
   \int_0^{2\pi} d\varphi \int_{-\infty}^{\infty} dz \int_0^{\infty} \rho d\rho  \Psi_{s, \ell}^{\uparrow , \downarrow}(x)^\dag \Psi_{s', \ell'}^{\uparrow , \downarrow}(x) \propto \delta_{s s'} \delta_{\ell \ell'}\delta(p_z-p_z').
\end{equation}
Importantly, $\Psi_{s, \ell}^{\uparrow , \downarrow}$ are the eigenfunctions of z-projection of the total angular momentum (TAM) operator
\begin{equation}
\begin{aligned}
    &\hat{j}_z \Psi_{s, \ell}^{\uparrow , \downarrow} = \left(\hat{\ell}_z + \frac{1}{2}\widehat{\Sigma}_z\right)  \Psi_{s, \ell}^{\uparrow , \downarrow} = \ell  \Psi_{s, \ell}^{\uparrow , \downarrow};\\
    & \widehat{\Sigma}_z = 
        \begin{pmatrix}
    \sigma_z & 0  \\
     0 & \sigma_z \\
         \end{pmatrix},
\end{aligned}
\end{equation}
but clearly not the eigenfunctions of $\hat{\ell}_z$ and $\widehat{\Sigma}_z$ separately. 

\subsection{Spin properties of the electron}
In a longitudinal magnetic field the  spin projection onto the field is not conserved  in the sense that the Pauli spin operator $\hat{S} = \frac{1}{2}\hat{\bm{\Sigma}}$ does not commute with the Dirac Hamiltonian. Many attempts have been made to introduce other spin operators that commute with the Hamiltonian and adequately describe the polarization of the particle \cite{ST, 7spin, Kudlis}. Since the TAM operator $\hat{\bm{j}} = \hat{\bm{r}} \times \hat{\bm{p}} + \frac{1}{2} \hat{\bm{\Sigma}}$ is well-defined for the Dirac equation and undisputed, different definitions of the spin operator imply different position operators, and hence different orbital angular momentum operators \cite{7spin, Bliohk2017_position}. Thus, any specific choice of the spin operator is equivalent to the splitting of the TAM into spin and orbital parts, which is ambiguous.

As the investigation of the polarization phenomena, such as well-studied Sokolov-Ternov effect \cite{ST, Kruining2019}, is not the purpose of this work, we do not follow the approach mentioned above. Alternatively, we claim that two types of states obtained in the previous section are sufficient to describe the states of the relativistic electron in magnetic field and derive the evolved state of the emitted photon because it is the total angular momentum projection $j_z$ of the electron that is observable in this problem and not $S_z$ alone.

Let us, however, evaluate the mean values of the spin operator $\hat{S}_z$ for the above states $\Psi_{s, \ell}^{\uparrow}$ and $\Psi_{s, \ell}^{\downarrow}$:
\begin{equation}
    S_z^{\uparrow , \downarrow} \equiv \frac{1}{2} \int d^3\bm{r}\: \left(\Psi_{s, \ell}^{\uparrow , \downarrow}{}\right)^\dag \hat{\Sigma}_z \Psi_{s, \ell}^{\uparrow , \downarrow} = \pm\frac{1}{2}\left(1-\frac{p_{\perp}^2}{\varepsilon(\varepsilon+m)}\right).
    \label{Sz1}
\end{equation}
Importantly, when the transverse momentum $p_{\perp}$ becomes non-relativistic, $p_{\perp} \ll m$, in the limit $H \ll H_c$, the spin projections tend to the values $\pm 1/2$. For this reason, we refer to the state $\Psi_{s, \ell}^{\uparrow}$ as ``spin-up'' and to $\Psi_{s, \ell}^{\downarrow}$ as ``spin-down''. 

\subsection{Alternative choice of the solutions}
\begin{figure}[h]
	\centering
	\begin{subfigure}{0.45\linewidth}
		\includegraphics[width=\linewidth]{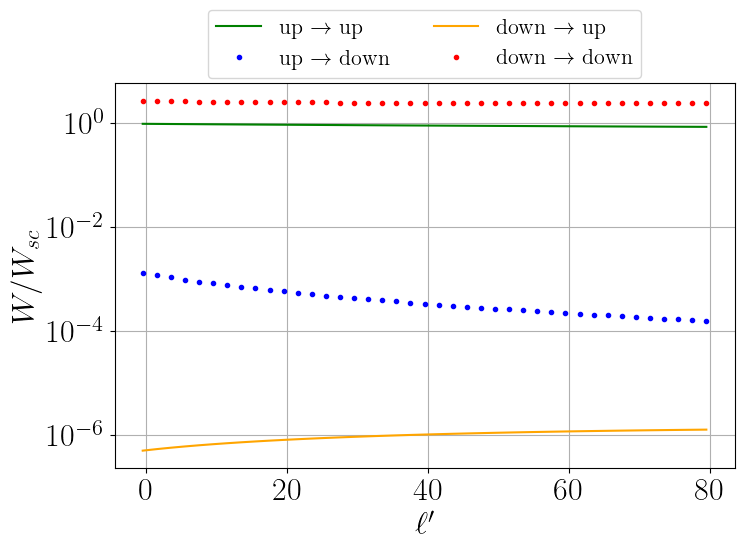}
	\end{subfigure}
	\begin{subfigure}{0.45\linewidth}
		\includegraphics[width=\linewidth]{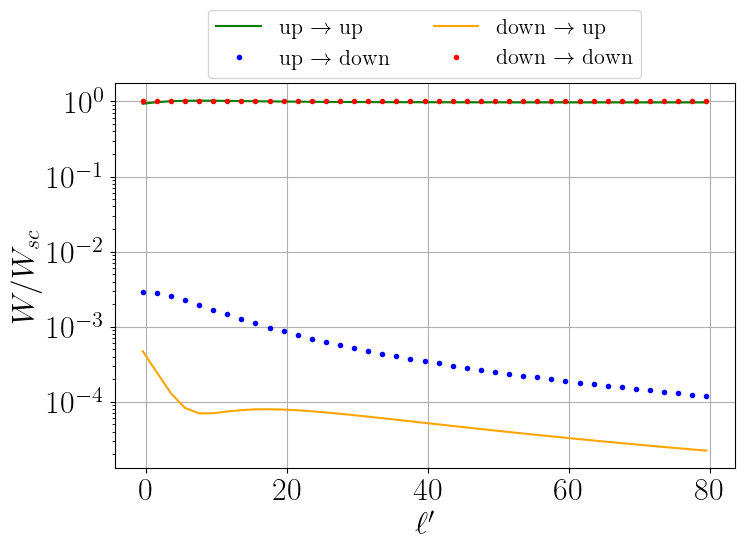}
	\end{subfigure}
 
	\caption{The ratio of four possible types of transition probabilities to the probability of emission by a scalar charge for the alternative choice of ``spin-up'' and ``spin-down'' states given by Eqs. \eqref{Psi_Plus_prime} and \eqref{Psi_Minus_prime}. $H=10^{-3}H_c$ (left) and $H=H_c$ (right); $\ell-\ell'=3$ for all transitions. The same as in Fig. \ref{ratios}, but for the alternative choice of ``spin-up'' and ``spin-down'' states given by Eqs. \eqref{Psi_Plus_prime} and \eqref{Psi_Minus_prime}.}
	\label{ratios_prime}
\end{figure}

\begin{figure}[h]
    \centering
    \includegraphics[width=\textwidth]{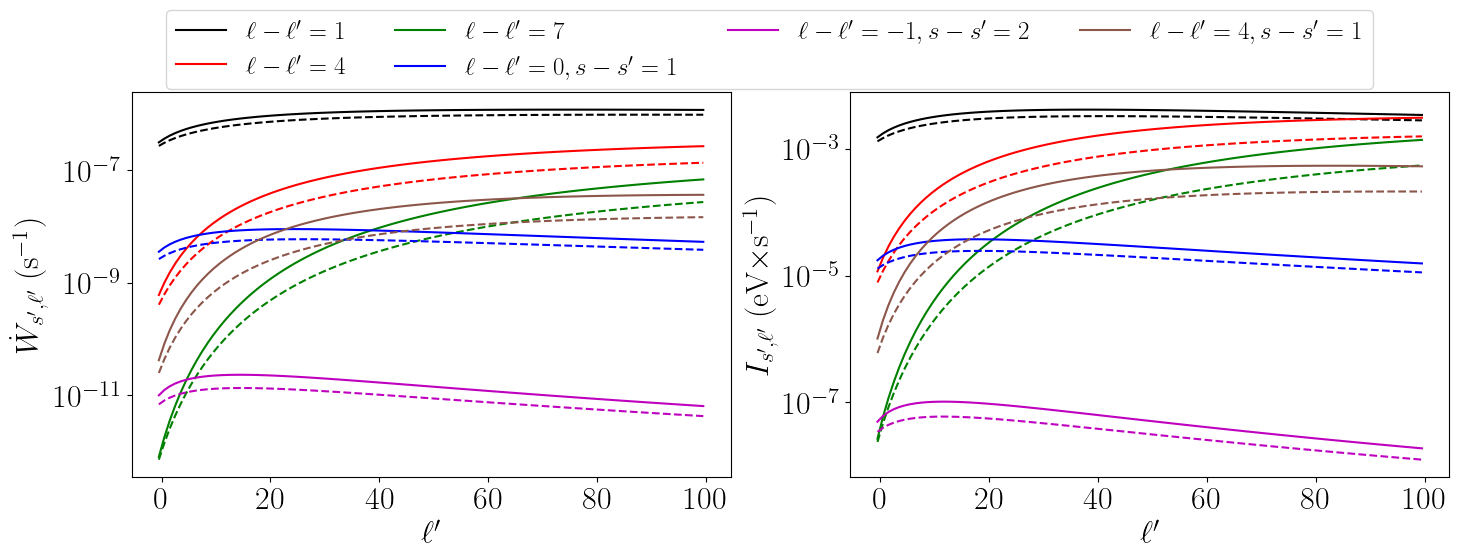}
     \caption{
     Comparison of the emission probability (left) and the corresponding intensity (right) for ``up-up'' transitions calculated for the ``spin-up'' states \eqref{Phi_Plus} (solid lines) and for the alternative states \eqref{Phi_Plus_prime} (dashed lines). Parameters: $H=0.01H_c$, $p_z=10^{-3} mc$ and $s = s' = 20$ unless stated otherwise.}
     \label{compare_W_I}
\end{figure}

The bispinors \eqref{Phi_Plus} and \eqref{Phi_Minus} are simple and intuitive, yet certainly not the only possible solutions of the Dirac equation \eqref{Eq:Dirac}. One can generally choose arbitrary constants $c_i$ in Eq. \eqref{sol_with_c_i} (see the discussion in \cite{ST}) but this arbitrariness does not affect the main property of the emitted photon: it represents a twisted Bessel beam with the TAM projection transferred from the initial electron, regardless of the values of $c_i$. The constants in Eq. \eqref{sol_with_c_i}, however, govern the spin properties of the states and the transition probabilities. Let us now consider another pair of ``auxillary'' solutions of \eqref{Eq:4KG} as, for instance,
\begin{equation}
\label{Phi_Plus_prime}
     \tilde{\Phi}_{s, \ell}^{\uparrow}(x) = \tilde{N}^{\uparrow}\Phi_{s, \ell - 1/2}(\rho)e^{-i t \varepsilon_{s, \ell} + i (\ell-1/2) \varphi+i p_z z} \begin{pmatrix}
        0\\
        0\\
        1\\
        0
    \end{pmatrix};
\end{equation}
\begin{equation}
\label{Phi_Minus_prime}
     \tilde{\Phi}_{s, \ell}^{\downarrow}(x) = \tilde{N}^{\downarrow}\Phi_{s, \ell + 1/2}(\rho)e^{-i t \varepsilon_{s, \ell} + i (\ell+1/2) \varphi+i p_z z} \begin{pmatrix}
        0\\
        0\\
        0\\
        1
    \end{pmatrix}.
\end{equation}
The corresponding ``projected'' solutions of the Dirac equation are
\begin{equation}
\label{Psi_Plus_prime}
    \tilde{\Psi}_{s, \ell}^{\uparrow}(x) = \tilde{N}^{\uparrow}
    \begin{pmatrix}
        -p_z \Phi_{s, \ell - 1/2}(\rho) e^{-i\varphi/2}\\
        ieH \Phi_{s, \ell + 1/2}(\rho)e^{i\varphi/2}\\
        (m-\varepsilon) \Phi_{s, \ell - 1/2}(\rho) e^{-i\varphi/2}\\
        0
    \end{pmatrix}
     e^{-i t \varepsilon_{s, \ell} + i \ell \varphi+i p_z z};
\end{equation}
\begin{equation} 
\label{Psi_Minus_prime}
    \tilde{\Psi}_{s, \ell}^{\downarrow}(x) = \tilde{N}^{\downarrow}
    \begin{pmatrix}
        2i(\ell+s+1/2) \Phi_{s, \ell - 1/2}(\rho) e^{-i\varphi/2}\\
         p_z \Phi_{s, \ell + 1/2}(\rho)e^{i\varphi/2}\\
       0\\
        (m-\varepsilon) \Phi_{s, \ell + 1/2}(\rho)e^{i\varphi/2}
    \end{pmatrix}
     e^{-i t \varepsilon_{s, \ell} + i \ell \varphi+i p_z z}.
\end{equation}
The mean value of the spin projection is (compare with Eq.(\ref{Sz1}))
\begin{equation}
    \tilde{S}_z^{\uparrow , \downarrow} = \frac{1}{2} \int d^3\bm{r}\: \left(\tilde{\Psi}_{s, \ell}^{\uparrow , \downarrow}\right)^\dag \hat{\Sigma}_z \tilde{\Psi}_{s, \ell}^{\uparrow , \downarrow} = \pm\frac{1}{2}\left(1-\frac{p_{\perp}^2}{\varepsilon(\varepsilon-m)}\right).
    \label{Sz2}
\end{equation}
For these states the orthogonality relations \eqref{orth_0} and \eqref{orth} are also satisfied and still $\hat{j}_z \tilde{\Psi}_{s, \ell}^{\uparrow , \downarrow} =  \ell  \tilde{\Psi}_{s,\ell}^{\uparrow , \downarrow}$. Although the probabilities and intensities shown in Fig. \ref{ratios_prime} differ from those in Fig. \ref{ratios} --- in the left figure the ratio of ``up-down'' transition probability to the ``down-up'' one is two orders of magnitude lower --- the self-polarization effect is present and qualitatively the picture is the same. Another comparison of the emission probabilities calculated for different polarization states is presented in Fig. \ref{compare_W_I}, demonstrating their qualitative agreement. This fact alongside with the orthogonality does not allow us to say that any of the two sets of states -- either $\{\Psi_{s, \ell}^{\uparrow},  \Psi_{s, \ell}^{\downarrow}\}$ or $\{\tilde{\Psi}_{s, \ell}^{\uparrow},  \tilde{\Psi}_{s, \ell}^{\downarrow}\}$ -- is more preferable for describing an electron in the magnetic field without discussing the spin detector properties. We stress, however, that these spin subtleties only affect the differential probability but not the total one and not the evolved state of the emitted photon, which is defined by the difference of the electron total angular momenta, $\ell-\ell'$.

Let us also discuss the auxiliary solutions which represent the sums of $\Phi_{s, \ell}^{\uparrow , \downarrow}$ and $\tilde{\Phi}_{s, \ell}^{\uparrow , \downarrow}$:
\begin{equation}
\label{Phi_Plus_prime_prime}
    \phi_{s, \ell}^{\uparrow}(x) = n^{\uparrow}\Phi_{s, \ell - 1/2}(\rho)e^{-i t \varepsilon_{s, \ell} + i (\ell-1/2) \varphi+i p_z z} \begin{pmatrix}
        1\\
        0\\
        1\\
        0
    \end{pmatrix};
\end{equation}
\begin{equation}
\label{Phi_Minus_prime_prime}
     \phi_{s, \ell}^{\downarrow}(x) = n^{\downarrow}\Phi_{s, \ell + 1/2}(\rho)e^{-i t \varepsilon_{s, \ell} + i (\ell+1/2) \varphi+i p_z z} \begin{pmatrix}
        0\\
        1\\
        0\\
        1
    \end{pmatrix}.
\end{equation}
The corresponding ``projected'' solutions of the Dirac equation are
\begin{equation}
\label{Psi_Plus_prime_prime}
    \psi_{s, \ell}^{\uparrow}(x) = n^{\uparrow}
    \begin{pmatrix}
        (m-p_z+\varepsilon) \Phi_{s, \ell - 1/2}(\rho) e^{-i\varphi/2}\\
        ieH \Phi_{s, \ell + 1/2}(\rho)e^{i\varphi/2}\\
        (m+p_z-\varepsilon) \Phi_{s, \ell - 1/2}(\rho) e^{-i\varphi/2}\\
        -ieH \Phi_{s, \ell + 1/2}(\rho)e^{i\varphi/2}
    \end{pmatrix}
     e^{-i t \varepsilon_{s, \ell} + i \ell \varphi+i p_z z};
\end{equation}
\begin{equation} 
\label{Psi_Minus_prime_prime}
    \psi_{s, \ell}^{\downarrow}(x) = n^{\downarrow}
    \begin{pmatrix}
        2i(\ell+s+1/2) \Phi_{s, \ell - 1/2}(\rho) e^{-i\varphi/2}\\
         (m+p_z+\varepsilon) \Phi_{s, \ell + 1/2}(\rho)e^{i\varphi/2}\\
        -2i(\ell+s+1/2)\Phi_{s, \ell - 1/2}(\rho) e^{-i\varphi/2}\\
        (m-p_z-\varepsilon) \Phi_{s, \ell + 1/2}(\rho)e^{i\varphi/2}
    \end{pmatrix}
     e^{-i t \varepsilon_{s, \ell} + i \ell \varphi+i p_z z}.
\end{equation}
These states, however, have certain shortcomings. First of all, $\psi_{s, \ell}^{\uparrow}$ and $\psi_{s, \ell}^{\downarrow}$ are not orthogonal:
\begin{equation}
   \int_0^{\infty} \rho d\rho \: \psi_{s, \ell}^{\uparrow}(x)^\dag \psi_{s, \ell}^{\downarrow}(x) \neq 0,
\end{equation}
which does not allow to speak about two distinct polarization states, yet the orthogonality with respect to $s$, $\ell$ and $p_z$ (See Eq. \eqref{orth}) still holds. Second, the 
dependence of the mean values of $\hat{s}_z$ on the sign of $p_z$ arises:

\begin{equation}
\label{s_z_prime_prime}
    s_z^{\uparrow , \downarrow} = \frac{1}{2} \int d^3\bm{r}\: \left(\psi_{s, \ell}^{\uparrow , \downarrow}\right)^\dag \hat{\Sigma}_z \psi_{s, \ell}^{\uparrow , \downarrow} = \pm\frac{1}{2}\left(1-\frac{p_{\perp}^2}{\varepsilon(\varepsilon\mp p_z)}\right).
\end{equation}

 If $p_z \gg \varepsilon$ or $p_z \ll \varepsilon$, Eq. \eqref{s_z_prime_prime} enables the effective spin flip, meaning that $s_z^{\uparrow} \to -\frac{1}{2}$ and $s_z^{\downarrow} \to \frac{1}{2}$, respectively, which means that the sign of $s_z$ is frame dependent and, thus, $s_z$ itself is not a good quantum number.

\begin{figure}[h]
	\centering
	\begin{subfigure}{0.45\linewidth}
		\includegraphics[width=\linewidth]{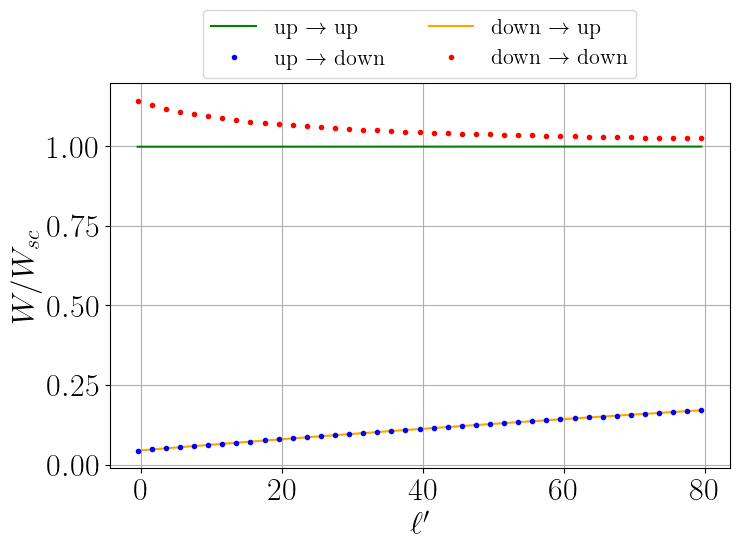}
	\end{subfigure}
	\begin{subfigure}{0.45\linewidth}
		\includegraphics[width=\linewidth]{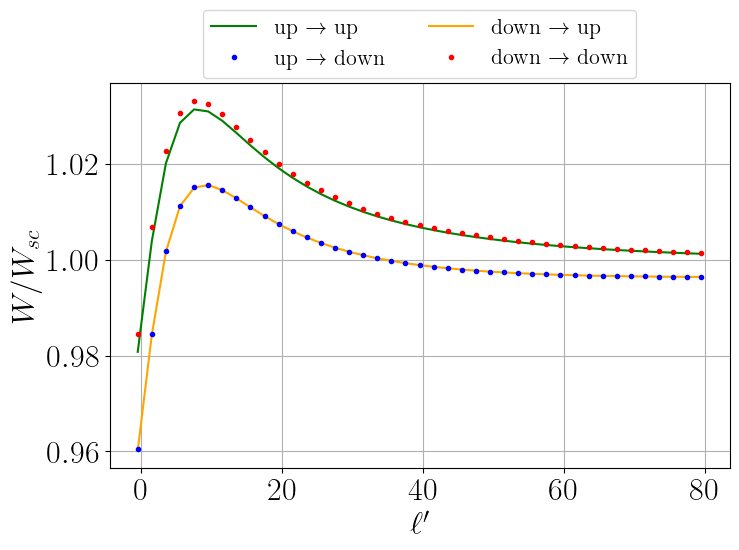}
	\end{subfigure}
 
	\caption{The same as in Figs. \ref{ratios} and \ref{ratios_prime}, but for the ``spin-up'' state \eqref{Psi_Plus_prime_prime} and the ``spin-down'' state \eqref{Psi_Minus_prime_prime}: ratio of four possible types of the transition probabilities to the probability of emission by a scalar charge. $H=10^{-3}H_c$ (left) and $H=H_c$ (right); $\ell-\ell'=3$, $s=s'=20$ for all transitions. In contrast to Figs. \ref{ratios} and \ref{ratios_prime}, the scale here is \textit{not logarithmic}.}
	\label{ratios_prime_prime}
\end{figure}

Finally, although the mean spin projections \eqref{s_z_prime_prime} tend to $\pm1/2$ for the weak field, these states do not predict the self-polarization effect even when $H \ll H_c$. Fig. \ref{ratios_prime_prime} demonstrates that the probabilities of ``up-down'' and ``down-up'' transitions are indistinguishable both in critical and subcritical magnetic field. Moreover, in a critical field the transition probabilities of all four types are close to those for a ``scalar electron'', which confirms that any effects associated with the spin alone are not adequately described with these states. It is also noteworthy that according to Figs. \ref{ratios}, \ref{ratios_prime} and \ref{ratios_prime_prime}, the probabilities of no spin-flip transitions are of the order of the probability for a scalar particle regardless of the spin states basis and the field strength.

\section{Appendix: Transition currents and amplitudes}
\label{Sec: Transition currents}

The transition current components for a transition between two ``spin-up'' states \eqref{Psi_i} and \eqref{Psi_f} are
\begin{multline}
\label{j0}
    j_{\uparrow \uparrow}^0(x) = N_i^{\uparrow} N_f^{\uparrow} \rho^{\ell+\ell'-1} e^{-2\rho^2/\rho_H^2 + i(p_z-p_z')z + i(\ell-\ell')\varphi - i(\varepsilon-\varepsilon')t} \times\\
   \times  \Bigg[\left(m(m+\varepsilon+\varepsilon')+p_zp_z'+\varepsilon\varepsilon'\right) L_s^{\ell-1/2}(2\rho^2/\rho_H^2) L_{s'}^{\ell'-1/2}(2\rho^2/\rho_H^2) + 4\rho^2/\rho_H^2 L_{s}^{\ell+1/2}(2\rho^2/\rho_H^2) L_{s'}^{\ell'+1/2}(2\rho^2/\rho_H^2) \Bigg];
\end{multline}

\begin{multline}
\label{j1}
   j_{\uparrow \uparrow}^1(x) = -ieH N_i^{\uparrow} N_f^{\uparrow} \rho^{\ell+\ell'} e^{-2\rho^2/\rho_H^2 + i(p_z-p_z')z + i(\ell-\ell')\varphi - i(\varepsilon-\varepsilon')t} \times\\
   \times  \Bigg[e^{i\varphi}\left(m+\varepsilon' \right) L_s^{\ell+1/2}(2\rho^2/\rho_H^2) L_{s'}^{\ell'-1/2}(2\rho^2/\rho_H^2) - e^{-i\varphi}(m+\varepsilon) L_{s}^{\ell-1/2}(2\rho^2/\rho_H^2) L_{s'}^{\ell'+1/2}(2\rho^2/\rho_H^2) \Bigg] ;
\end{multline}

\begin{multline}
\label{j2}
   j_{\uparrow \uparrow}^2(x) = -eH N_i^{\uparrow} N_f^{\uparrow} \rho^{\ell+\ell'} e^{-2\rho^2/\rho_H^2 + i(p_z-p_z')z + i(\ell-\ell')\varphi - i(\varepsilon-\varepsilon')t} \times\\
   \times  \Bigg[e^{i\varphi}\left(m+\varepsilon' \right) L_s^{\ell+1/2}(2\rho^2/\rho_H^2) L_{s'}^{\ell'-1/2}(2\rho^2/\rho_H^2) + e^{-i\varphi}(m+\varepsilon) L_{s}^{\ell-1/2}(2\rho^2/\rho_H^2) L_{s'}^{\ell'+1/2}(2\rho^2/\rho_H^2) \Bigg] ;
\end{multline}

\begin{multline}
\label{j3}
   j_{\uparrow \uparrow}^3(x) = 2 N_i^{\uparrow} N_f^{\uparrow} \rho^{\ell+\ell'-1} e^{-2\rho^2/\rho_H^2 + i(p_z-p_z')z + i(\ell-\ell')\varphi - i(\varepsilon-\varepsilon')t} \times\\
   \times \big(m(p_z+p_z') +p_z'\varepsilon + p_z\varepsilon'\big) L_s^{\ell-1/2}(2\rho^2/\rho_H^2) L_{s'}^{\ell'-1/2}(2\rho^2/\rho_H^2).
\end{multline}

To calculate the S-matrix amplitude we first take the integrals over $t$, $z$ and $\varphi$ in Eq. \eqref{S_fi} and leave the radial part unevaluated. The azimuthal integral can be taken by employing the integral representation of the Bessel functions

\begin{equation}
    \int_0^{2\pi} \frac{d \varphi}{2\pi} e^{i \ell \varphi + i x \cos \varphi} = i^{\ell} J_{\ell}(x).
\end{equation}
Thus, we arrive at the following expression
\begin{equation}
    S_{\uparrow \uparrow}^{(1)} = \int_0^{\infty} \tilde{S}_{\uparrow \uparrow}^{(1)} d \rho,
\end{equation}
where the integrand is
\begin{multline}
    \tilde{S}_{\uparrow \uparrow}^{(1)} =  ei^{-\ell+\ell'+1}(2\pi)^3\frac{N_i^\uparrow N_f^\uparrow}{\sqrt{2\omega V}}\delta(\omega + \varepsilon' - \varepsilon) \delta(k_z + p_z' - p_z) e^{i(\ell-\ell')\varphi_k-2\tilde{\rho}^2} \times\\
    \times\Bigg[\big(m(p_z+p_z')+p_z'\varepsilon+p_z\varepsilon' \big) d_{0\lambda}^{(1)}(\theta) \tilde{\rho}^{\ell+\ell'}\rho_H^{\ell+\ell'}J_{\ell-\ell'}(y\tilde{\rho})L_s^{\ell-1/2}(2\tilde{\rho}^2) L_{s'}^{\ell'-1/2}(2\tilde{\rho}^2)  -\\
    \sqrt{2}eH(m+\varepsilon') d_{-1\lambda}^{(1)} (\theta)\tilde{\rho}^{\ell+\ell'+1}\rho_H^{\ell+\ell'+1}J_{\ell-\ell'+1}(y\tilde{\rho})L_s^{\ell+1/2}(2\tilde{\rho}^2) L_{s'}^{\ell'-1/2}(2\tilde{\rho}^2) +\\
    \sqrt{2}eH(m+\varepsilon) d_{1\lambda}^{(1)} (\theta)\tilde{\rho}^{\ell+\ell'+1}\rho_H^{\ell+\ell'+1}J_{\ell-\ell'-1}(y\tilde{\rho})L_s^{\ell-1/2}(2\tilde{\rho}^2) L_{s'}^{\ell'+1/2}(2\tilde{\rho}^2) \Bigg].
\end{multline}
Here we have denoted $\tilde{\rho} \equiv \rho / \rho_H$ and $y \equiv k_{\perp} \rho_H$.
Then we use the following integral from \cite{Gr-R} (there is, however, a misprint $m\leftrightarrow n$)
\begin{multline}
\label{eq:F}
    \int\limits_0^{\infty} dx\,x^{\ell+\ell'+1}\, L_{s}^{\ell}(2x^2)L_{s'}^{\ell'}(2x^2)\,J_{\ell-\ell'}(y x)\, e^{-2x^2}  =\\
    \frac{(s'+\ell')!}{s!}\frac{1}{2^{3(s-s') + 2\ell - \ell' + 2}}\, y^{2(s-s') + \ell-\ell'}\, L_{s'+\ell'}^{s-s'+\ell-\ell'}\left(y^2/8\right)L_{s'}^{s-s'}\left(y^2/8\right)\, e^{-y^2/8} \equiv F^{\ell, \ell'}_{s, s'}(y)
\end{multline}
to obtain the expression \eqref{S_uu}.

The S-matrix elements for three other types of transitions are
\begin{multline}
\label{S_ud}
   S_{\uparrow \downarrow}^{(1)} =  e i^{-\ell+\ell'}(2\pi)^3\frac{N_i^{\uparrow} N_f^{\downarrow}}{\sqrt{2\omega V}}\delta(\omega + \varepsilon' - \varepsilon) \delta(k_z + p_z' - p_z) e^{i(\ell-\ell')\varphi_k} \times\\
    \times\Bigg[d_{0\lambda}^{(1)}(\theta) \rho_H^{\ell+\ell'+1}\left[-\left(m+\varepsilon\right)\left(1+2s'+2\ell'\right)   F_{s, s'}^{\ell-1/2, \ell'-1/2}(y) + 4 (m+\varepsilon') F_{s, s'}^{\ell+1/2, \ell'+1/2}(y)\right] -\\
    -\sqrt{2}\rho_H^{\ell+\ell'+2}\big( m(p_z'-p_z)+p_z'\varepsilon-p_z\varepsilon'\big) d_{1\lambda}^{(1)} (\theta)F_{s, s'}^{\ell-1/2, \ell'+1/2}(y)  \Bigg];
\end{multline}

\begin{multline}
\label{S_du}
    S_{\downarrow \uparrow}^{(1)} =  e i^{-\ell+\ell'}(2\pi)^3\frac{N_i^{\downarrow} N_f^{\uparrow}}{\sqrt{2\omega V}}\delta(\omega + \varepsilon' - \varepsilon) \delta(k_z + p_z' - p_z) e^{i(\ell-\ell')\varphi_k} \times\\
    \times\Bigg[d_{0\lambda}^{(1)}(\theta) \rho_H^{\ell+\ell'+1}\left[-\left(m+\varepsilon'\right)\left(1+2s+2\ell\right)   F_{s, s'}^{\ell-1/2, \ell'-1/2}(y) + 4 (m+\varepsilon) F_{s, s'}^{\ell+1/2, \ell'+1/2}(y)\right] -\\
    -\sqrt{2}\rho_H^{\ell+\ell'+2}\big(m(p_z'-p_z)+\varepsilon p_z'-\varepsilon'p_z\big) d_{-1\lambda}^{(1)}(\theta)  F_{s, s'}^{\ell+1/2, \ell'-1/2}(y)  \Bigg];
\end{multline}

\begin{multline}
\label{S_dd}
    S_{\downarrow \downarrow}^{(1)} =  e i^{-\ell+\ell'+1}(2\pi)^3\frac{N_i^{\downarrow} N_f^{\downarrow}}{\sqrt{2\omega V}}\delta(\omega + \varepsilon' - \varepsilon) \delta(k_z + p_z' - p_z) e^{i(\ell-\ell')\varphi_k} \times\\
    \times\Bigg[d_{0\lambda}^{(1)}(\theta) \rho_H^{\ell+\ell'+3}\big(m(p_z+p_z')+p_z'\varepsilon+p_z\varepsilon' \big) F_{s, s'}^{\ell+1/2, \ell'+1/2}(y)-\\
    \sqrt{2}\rho_H^{\ell+\ell'+2}\left[\left(1+2s'+2\ell'\right)(m+\varepsilon) d_{-1\lambda}^{(1)}(\theta)  F_{s, s'}^{\ell+1/2, \ell'-1/2}(y) - \left(1+2s+2\ell\right)(m+\varepsilon') d_{1\lambda}^{(1)} (\theta)F_{s, s'}^{\ell-1/2, \ell'+1/2}(y)  \right]\Bigg].
\end{multline}


\bibliographystyle{apsrev}
\bibliography{references}

\begin{thebibliography}{35}
\expandafter\ifx\csname natexlab\endcsname\relax\def\natexlab#1{#1}\fi
\expandafter\ifx\csname bibnamefont\endcsname\relax
  \def\bibnamefont#1{#1}\fi
\expandafter\ifx\csname bibfnamefont\endcsname\relax
  \def\bibfnamefont#1{#1}\fi
\expandafter\ifx\csname citenamefont\endcsname\relax
  \def\citenamefont#1{#1}\fi
\expandafter\ifx\csname url\endcsname\relax
  \def\url#1{\texttt{#1}}\fi
\expandafter\ifx\csname urlprefix\endcsname\relax\def\urlprefix{URL }\fi
\providecommand{\bibinfo}[2]{#2}
\providecommand{\eprint}[2][]{\url{#2}}

\bibitem[{\citenamefont{Katoh et~al.}(2017{\natexlab{a}})\citenamefont{Katoh,
  Fujimoto, Mirian, Konomi, Taira, Kaneyasu, Hosaka, Yamamoto, Mochihashi,
  Takashima et~al.}}]{katoh2017helical}
\bibinfo{author}{\bibfnamefont{M.}~\bibnamefont{Katoh}},
  \bibinfo{author}{\bibfnamefont{M.}~\bibnamefont{Fujimoto}},
  \bibinfo{author}{\bibfnamefont{N.}~\bibnamefont{Mirian}},
  \bibinfo{author}{\bibfnamefont{T.}~\bibnamefont{Konomi}},
  \bibinfo{author}{\bibfnamefont{Y.}~\bibnamefont{Taira}},
  \bibinfo{author}{\bibfnamefont{T.}~\bibnamefont{Kaneyasu}},
  \bibinfo{author}{\bibfnamefont{M.}~\bibnamefont{Hosaka}},
  \bibinfo{author}{\bibfnamefont{N.}~\bibnamefont{Yamamoto}},
  \bibinfo{author}{\bibfnamefont{A.}~\bibnamefont{Mochihashi}},
  \bibinfo{author}{\bibfnamefont{Y.}~\bibnamefont{Takashima}},
  \bibnamefont{et~al.}, \bibinfo{journal}{Scientific Reports}
  \textbf{\bibinfo{volume}{7}}, \bibinfo{pages}{6130}
  (\bibinfo{year}{2017}{\natexlab{a}}).

\bibitem[{\citenamefont{Katoh et~al.}(2017{\natexlab{b}})\citenamefont{Katoh,
  Fujimoto, Kawaguchi, Tsuchiya, Ohmi, Kaneyasu, Taira, Hosaka, Mochihashi, and
  Takashima}}]{katoh2017angular}
\bibinfo{author}{\bibfnamefont{M.}~\bibnamefont{Katoh}},
  \bibinfo{author}{\bibfnamefont{M.}~\bibnamefont{Fujimoto}},
  \bibinfo{author}{\bibfnamefont{H.}~\bibnamefont{Kawaguchi}},
  \bibinfo{author}{\bibfnamefont{K.}~\bibnamefont{Tsuchiya}},
  \bibinfo{author}{\bibfnamefont{K.}~\bibnamefont{Ohmi}},
  \bibinfo{author}{\bibfnamefont{T.}~\bibnamefont{Kaneyasu}},
  \bibinfo{author}{\bibfnamefont{Y.}~\bibnamefont{Taira}},
  \bibinfo{author}{\bibfnamefont{M.}~\bibnamefont{Hosaka}},
  \bibinfo{author}{\bibfnamefont{A.}~\bibnamefont{Mochihashi}},
  \bibnamefont{and}
  \bibinfo{author}{\bibfnamefont{Y.}~\bibnamefont{Takashima}},
  \bibinfo{journal}{Physical Review Letters} \textbf{\bibinfo{volume}{118}},
  \bibinfo{pages}{094801} (\bibinfo{year}{2017}{\natexlab{b}}).

\bibitem[{\citenamefont{Epp and Guselnikova}(2019)}]{epp2019angular}
\bibinfo{author}{\bibfnamefont{V.}~\bibnamefont{Epp}} \bibnamefont{and}
  \bibinfo{author}{\bibfnamefont{U.}~\bibnamefont{Guselnikova}},
  \bibinfo{journal}{Physics Letters A} \textbf{\bibinfo{volume}{383}},
  \bibinfo{pages}{2668} (\bibinfo{year}{2019}).

\bibitem[{\citenamefont{Epp et~al.}(2022)\citenamefont{Epp, Guselnikova, and
  Kamenskaya}}]{epp2022angular}
\bibinfo{author}{\bibfnamefont{V.}~\bibnamefont{Epp}},
  \bibinfo{author}{\bibfnamefont{U.}~\bibnamefont{Guselnikova}},
  \bibnamefont{and}
  \bibinfo{author}{\bibfnamefont{I.}~\bibnamefont{Kamenskaya}},
  \bibinfo{journal}{Physical Review A} \textbf{\bibinfo{volume}{105}},
  \bibinfo{pages}{023511} (\bibinfo{year}{2022}).

\bibitem[{\citenamefont{Allen et~al.}(1992)\citenamefont{Allen, Beijersbergen,
  Spreeuw, and Woerdman}}]{allen1992orbital}
\bibinfo{author}{\bibfnamefont{L.}~\bibnamefont{Allen}},
  \bibinfo{author}{\bibfnamefont{M.~W.} \bibnamefont{Beijersbergen}},
  \bibinfo{author}{\bibfnamefont{R.}~\bibnamefont{Spreeuw}}, \bibnamefont{and}
  \bibinfo{author}{\bibfnamefont{J.}~\bibnamefont{Woerdman}},
  \bibinfo{journal}{Physical review A} \textbf{\bibinfo{volume}{45}},
  \bibinfo{pages}{8185} (\bibinfo{year}{1992}).

\bibitem[{\citenamefont{Andrews and Babiker}(2012)}]{andrews2012angular}
\bibinfo{author}{\bibfnamefont{D.~L.} \bibnamefont{Andrews}} \bibnamefont{and}
  \bibinfo{author}{\bibfnamefont{M.}~\bibnamefont{Babiker}},
  \emph{\bibinfo{title}{The angular momentum of light}}
  (\bibinfo{publisher}{Cambridge University Press}, \bibinfo{year}{2012}).

\bibitem[{\citenamefont{Torres and Torner}(2011)}]{torres2011twisted}
\bibinfo{author}{\bibfnamefont{J.~P.} \bibnamefont{Torres}} \bibnamefont{and}
  \bibinfo{author}{\bibfnamefont{L.}~\bibnamefont{Torner}},
  \emph{\bibinfo{title}{Twisted photons: applications of light with orbital
  angular momentum}} (\bibinfo{publisher}{John Wiley \& Sons},
  \bibinfo{year}{2011}).

\bibitem[{\citenamefont{Knyazev and Serbo}(2018)}]{SerboUFN}
\bibinfo{author}{\bibfnamefont{B.~A.} \bibnamefont{Knyazev}} \bibnamefont{and}
  \bibinfo{author}{\bibfnamefont{V.~G.} \bibnamefont{Serbo}},
  \bibinfo{journal}{Physics-Uspekhi} \textbf{\bibinfo{volume}{61}},
  \bibinfo{pages}{449} (\bibinfo{year}{2018}),
  \urlprefix\url{https://dx.doi.org/10.3367/UFNe.2018.02.038306}.

\bibitem[{\citenamefont{Karlovets et~al.}(2023)\citenamefont{Karlovets,
  Baturin, Geloni, Sizykh, and Serbo}}]{EPJC}
\bibinfo{author}{\bibfnamefont{D.~V.} \bibnamefont{Karlovets}},
  \bibinfo{author}{\bibfnamefont{S.~S.} \bibnamefont{Baturin}},
  \bibinfo{author}{\bibfnamefont{G.}~\bibnamefont{Geloni}},
  \bibinfo{author}{\bibfnamefont{G.~K.} \bibnamefont{Sizykh}},
  \bibnamefont{and} \bibinfo{author}{\bibfnamefont{V.~G.} \bibnamefont{Serbo}},
  \bibinfo{journal}{The European Physical Journal C}
  \textbf{\bibinfo{volume}{83}}, \bibinfo{pages}{372} (\bibinfo{year}{2023}),
  ISSN \bibinfo{issn}{1434-6052},
  \urlprefix\url{https://doi.org/10.1140/epjc/s10052-023-11529-4}.

\bibitem[{\citenamefont{Sokolov and Ternov}(1974)}]{ST}
\bibinfo{author}{\bibfnamefont{A.}~\bibnamefont{Sokolov}} \bibnamefont{and}
  \bibinfo{author}{\bibfnamefont{I.}~\bibnamefont{Ternov}},
  \emph{\bibinfo{title}{Relativistic electron}} (\bibinfo{publisher}{Moscow
  Izdatel Nauka}, \bibinfo{year}{1974}).

\bibitem[{\citenamefont{Bordovitsyn et~al.}(1999)}]{bordovitsyn1999synchrotron}
\bibinfo{author}{\bibfnamefont{V.~A.} \bibnamefont{Bordovitsyn}}
  \bibnamefont{et~al.}, \emph{\bibinfo{title}{Synchrotron radiation theory and
  its development: in memory of IM Ternov}}, vol.~\bibinfo{volume}{5}
  (\bibinfo{publisher}{World Scientific}, \bibinfo{year}{1999}).

\bibitem[{\citenamefont{van Kruining et~al.}(2019)\citenamefont{van Kruining,
  Mackenroth, and G\"otte}}]{Kruining2019}
\bibinfo{author}{\bibfnamefont{K.}~\bibnamefont{van Kruining}},
  \bibinfo{author}{\bibfnamefont{F.}~\bibnamefont{Mackenroth}},
  \bibnamefont{and} \bibinfo{author}{\bibfnamefont{J.~B.}
  \bibnamefont{G\"otte}}, \bibinfo{journal}{Phys. Rev. D}
  \textbf{\bibinfo{volume}{100}}, \bibinfo{pages}{056014}
  (\bibinfo{year}{2019}),
  \urlprefix\url{https://link.aps.org/doi/10.1103/PhysRevD.100.056014}.

\bibitem[{\citenamefont{Maruyama et~al.}(2022)\citenamefont{Maruyama, Hayakawa,
  Kajino, and Cheoun}}]{Maruyama}
\bibinfo{author}{\bibfnamefont{T.}~\bibnamefont{Maruyama}},
  \bibinfo{author}{\bibfnamefont{T.}~\bibnamefont{Hayakawa}},
  \bibinfo{author}{\bibfnamefont{T.}~\bibnamefont{Kajino}}, \bibnamefont{and}
  \bibinfo{author}{\bibfnamefont{M.-K.} \bibnamefont{Cheoun}},
  \bibinfo{journal}{Physics Letters B} \textbf{\bibinfo{volume}{826}},
  \bibinfo{pages}{136779} (\bibinfo{year}{2022}), ISSN
  \bibinfo{issn}{0370-2693},
  \urlprefix\url{https://www.sciencedirect.com/science/article/pii/S037026932100719X}.

\bibitem[{\citenamefont{Zhang et~al.}(2020)\citenamefont{Zhang, Xu, and
  Jiang}}]{Zhang2020}
\bibinfo{author}{\bibfnamefont{C.}~\bibnamefont{Zhang}},
  \bibinfo{author}{\bibfnamefont{P.}~\bibnamefont{Xu}}, \bibnamefont{and}
  \bibinfo{author}{\bibfnamefont{X.}~\bibnamefont{Jiang}},
  \bibinfo{journal}{AIP Advances} \textbf{\bibinfo{volume}{10}},
  \bibinfo{pages}{105230} (\bibinfo{year}{2020}), ISSN
  \bibinfo{issn}{2158-3226},
  \eprint{https://pubs.aip.org/aip/adv/article-pdf/doi/10.1063/5.0019899/8399921/105230\_1\_online.pdf},
  \urlprefix\url{https://doi.org/10.1063/5.0019899}.

\bibitem[{\citenamefont{Karlovets and Di~Piazza}(2023)}]{Karlovets2023}
\bibinfo{author}{\bibfnamefont{D.}~\bibnamefont{Karlovets}} \bibnamefont{and}
  \bibinfo{author}{\bibfnamefont{A.}~\bibnamefont{Di~Piazza}},
  \bibinfo{journal}{Phys. Rev. D} \textbf{\bibinfo{volume}{108}},
  \bibinfo{pages}{063007} (\bibinfo{year}{2023}),
  \urlprefix\url{https://link.aps.org/doi/10.1103/PhysRevD.108.063007}.

\bibitem[{\citenamefont{Sadowski et~al.}(2006)\citenamefont{Sadowski, Martinez,
  Potemski, Berger, and de~Heer}}]{Sadovsky}
\bibinfo{author}{\bibfnamefont{M.~L.} \bibnamefont{Sadowski}},
  \bibinfo{author}{\bibfnamefont{G.}~\bibnamefont{Martinez}},
  \bibinfo{author}{\bibfnamefont{M.}~\bibnamefont{Potemski}},
  \bibinfo{author}{\bibfnamefont{C.}~\bibnamefont{Berger}}, \bibnamefont{and}
  \bibinfo{author}{\bibfnamefont{W.~A.} \bibnamefont{de~Heer}},
  \bibinfo{journal}{Phys. Rev. Lett.} \textbf{\bibinfo{volume}{97}},
  \bibinfo{pages}{266405} (\bibinfo{year}{2006}),
  \urlprefix\url{https://link.aps.org/doi/10.1103/PhysRevLett.97.266405}.

\bibitem[{\citenamefont{Li and Andrei}(2007)}]{Li}
\bibinfo{author}{\bibfnamefont{G.}~\bibnamefont{Li}} \bibnamefont{and}
  \bibinfo{author}{\bibfnamefont{E.~Y.} \bibnamefont{Andrei}},
  \bibinfo{journal}{Nature physics} \textbf{\bibinfo{volume}{3}},
  \bibinfo{pages}{623} (\bibinfo{year}{2007}).

\bibitem[{\citenamefont{Yang et~al.}(2010)\citenamefont{Yang, Peeters, and
  Xu}}]{Yang}
\bibinfo{author}{\bibfnamefont{C.~H.} \bibnamefont{Yang}},
  \bibinfo{author}{\bibfnamefont{F.~M.} \bibnamefont{Peeters}},
  \bibnamefont{and} \bibinfo{author}{\bibfnamefont{W.}~\bibnamefont{Xu}},
  \bibinfo{journal}{Phys. Rev. B} \textbf{\bibinfo{volume}{82}},
  \bibinfo{pages}{075401} (\bibinfo{year}{2010}),
  \urlprefix\url{https://link.aps.org/doi/10.1103/PhysRevB.82.075401}.

\bibitem[{\citenamefont{Zhang et~al.}(2006)\citenamefont{Zhang, Jiang, Small,
  Purewal, Tan, Fazlollahi, Chudow, Jaszczak, Stormer, and Kim}}]{LLsplitting}
\bibinfo{author}{\bibfnamefont{Y.}~\bibnamefont{Zhang}},
  \bibinfo{author}{\bibfnamefont{Z.}~\bibnamefont{Jiang}},
  \bibinfo{author}{\bibfnamefont{J.~P.} \bibnamefont{Small}},
  \bibinfo{author}{\bibfnamefont{M.~S.} \bibnamefont{Purewal}},
  \bibinfo{author}{\bibfnamefont{Y.-W.} \bibnamefont{Tan}},
  \bibinfo{author}{\bibfnamefont{M.}~\bibnamefont{Fazlollahi}},
  \bibinfo{author}{\bibfnamefont{J.~D.} \bibnamefont{Chudow}},
  \bibinfo{author}{\bibfnamefont{J.~A.} \bibnamefont{Jaszczak}},
  \bibinfo{author}{\bibfnamefont{H.~L.} \bibnamefont{Stormer}},
  \bibnamefont{and} \bibinfo{author}{\bibfnamefont{P.}~\bibnamefont{Kim}},
  \bibinfo{journal}{Phys. Rev. Lett.} \textbf{\bibinfo{volume}{96}},
  \bibinfo{pages}{136806} (\bibinfo{year}{2006}),
  \urlprefix\url{https://link.aps.org/doi/10.1103/PhysRevLett.96.136806}.

\bibitem[{\citenamefont{Shah et~al.}(2012)\citenamefont{Shah, Iqbal,
  Tsintsadze, Masood, and Qureshi}}]{Shah}
\bibinfo{author}{\bibfnamefont{H.~A.} \bibnamefont{Shah}},
  \bibinfo{author}{\bibfnamefont{M.~J.} \bibnamefont{Iqbal}},
  \bibinfo{author}{\bibfnamefont{N.}~\bibnamefont{Tsintsadze}},
  \bibinfo{author}{\bibfnamefont{W.}~\bibnamefont{Masood}}, \bibnamefont{and}
  \bibinfo{author}{\bibfnamefont{M.~N.~S.} \bibnamefont{Qureshi}},
  \bibinfo{journal}{Physics of Plasmas} \textbf{\bibinfo{volume}{19}},
  \bibinfo{pages}{092304} (\bibinfo{year}{2012}), ISSN
  \bibinfo{issn}{1070-664X},
  \eprint{https://pubs.aip.org/aip/pop/article-pdf/doi/10.1063/1.4752416/16125112/092304\_1\_online.pdf},
  \urlprefix\url{https://doi.org/10.1063/1.4752416}.

\bibitem[{\citenamefont{Larson et~al.}(1988)\citenamefont{Larson, Edge,
  Elmquist, Mansour, and Trainham}}]{Larson_1988}
\bibinfo{author}{\bibfnamefont{D.~J.} \bibnamefont{Larson}},
  \bibinfo{author}{\bibfnamefont{C.~J.} \bibnamefont{Edge}},
  \bibinfo{author}{\bibfnamefont{R.~E.} \bibnamefont{Elmquist}},
  \bibinfo{author}{\bibfnamefont{N.~B.} \bibnamefont{Mansour}},
  \bibnamefont{and} \bibinfo{author}{\bibfnamefont{R.}~\bibnamefont{Trainham}},
  \bibinfo{journal}{Physica Scripta} \textbf{\bibinfo{volume}{1988}},
  \bibinfo{pages}{183} (\bibinfo{year}{1988}),
  \urlprefix\url{https://dx.doi.org/10.1088/0031-8949/1988/T22/028}.

\bibitem[{\citenamefont{Kluge et~al.}(2003)\citenamefont{Kluge, Blaum,
  Herfurth, and Quint}}]{Kluge_2003}
\bibinfo{author}{\bibfnamefont{H.-J.} \bibnamefont{Kluge}},
  \bibinfo{author}{\bibfnamefont{K.}~\bibnamefont{Blaum}},
  \bibinfo{author}{\bibfnamefont{F.}~\bibnamefont{Herfurth}}, \bibnamefont{and}
  \bibinfo{author}{\bibfnamefont{W.}~\bibnamefont{Quint}},
  \bibinfo{journal}{Physica Scripta} \textbf{\bibinfo{volume}{2003}},
  \bibinfo{pages}{167} (\bibinfo{year}{2003}),
  \urlprefix\url{https://dx.doi.org/10.1238/Physica.Topical.104a00167}.

\bibitem[{\citenamefont{Bagrov and Gitman}(2014)}]{bagrov2014dirac}
\bibinfo{author}{\bibfnamefont{V.~G.} \bibnamefont{Bagrov}} \bibnamefont{and}
  \bibinfo{author}{\bibfnamefont{D.}~\bibnamefont{Gitman}},
  \emph{\bibinfo{title}{The Dirac equation and its solutions}},
  vol.~\bibinfo{volume}{4} (\bibinfo{publisher}{Walter de Gruyter GmbH \& Co
  KG}, \bibinfo{year}{2014}).

\bibitem[{\citenamefont{Bogdanov et~al.}(2018)\citenamefont{Bogdanov, Kazinski,
  and Lazarenko}}]{Bogdanov2018}
\bibinfo{author}{\bibfnamefont{O.~V.} \bibnamefont{Bogdanov}},
  \bibinfo{author}{\bibfnamefont{P.~O.} \bibnamefont{Kazinski}},
  \bibnamefont{and} \bibinfo{author}{\bibfnamefont{G.~Y.}
  \bibnamefont{Lazarenko}}, \bibinfo{journal}{Phys. Rev. A}
  \textbf{\bibinfo{volume}{97}}, \bibinfo{pages}{033837}
  (\bibinfo{year}{2018}),
  \urlprefix\url{https://link.aps.org/doi/10.1103/PhysRevA.97.033837}.

\bibitem[{\citenamefont{Bogdanov et~al.}(2019)\citenamefont{Bogdanov, Kazinski,
  and Lazarenko}}]{Bogdanov2019}
\bibinfo{author}{\bibfnamefont{O.~V.} \bibnamefont{Bogdanov}},
  \bibinfo{author}{\bibfnamefont{P.~O.} \bibnamefont{Kazinski}},
  \bibnamefont{and} \bibinfo{author}{\bibfnamefont{G.~Y.}
  \bibnamefont{Lazarenko}}, \bibinfo{journal}{Phys. Rev. D}
  \textbf{\bibinfo{volume}{99}}, \bibinfo{pages}{116016}
  (\bibinfo{year}{2019}),
  \urlprefix\url{https://link.aps.org/doi/10.1103/PhysRevD.99.116016}.

\bibitem[{\citenamefont{Berestetskii et~al.}(1982)\citenamefont{Berestetskii,
  Lifshitz, and Pitaevskii}}]{BLP}
\bibinfo{author}{\bibfnamefont{V.~B.} \bibnamefont{Berestetskii}},
  \bibinfo{author}{\bibfnamefont{E.~M.} \bibnamefont{Lifshitz}},
  \bibnamefont{and} \bibinfo{author}{\bibfnamefont{L.~P.}
  \bibnamefont{Pitaevskii}}, \emph{\bibinfo{title}{Quantum Electrodynamics:
  Volume 4}}, vol.~\bibinfo{volume}{4}
  (\bibinfo{publisher}{Butterworth-Heinemann}, \bibinfo{year}{1982}).

\bibitem[{\citenamefont{Scully and Zubairy}(1999)}]{Scully}
\bibinfo{author}{\bibfnamefont{M.~O.} \bibnamefont{Scully}} \bibnamefont{and}
  \bibinfo{author}{\bibfnamefont{M.~S.} \bibnamefont{Zubairy}},
  \emph{\bibinfo{title}{Quantum optics}} (\bibinfo{year}{1999}).

\bibitem[{\citenamefont{Varshalovich et~al.}(1988)\citenamefont{Varshalovich,
  Moskalev, and Khersonskii}}]{Varshalovich}
\bibinfo{author}{\bibfnamefont{D.~A.} \bibnamefont{Varshalovich}},
  \bibinfo{author}{\bibfnamefont{A.~N.} \bibnamefont{Moskalev}},
  \bibnamefont{and} \bibinfo{author}{\bibfnamefont{V.~K.}
  \bibnamefont{Khersonskii}}, \emph{\bibinfo{title}{Quantum Theory of Angular
  Momentum}} (\bibinfo{year}{1988}),
  \urlprefix\url{https://api.semanticscholar.org/CorpusID:117798939}.

\bibitem[{\citenamefont{Bauke et~al.}(2014)\citenamefont{Bauke, Ahrens, Keitel,
  and Grobe}}]{7spin}
\bibinfo{author}{\bibfnamefont{H.}~\bibnamefont{Bauke}},
  \bibinfo{author}{\bibfnamefont{S.}~\bibnamefont{Ahrens}},
  \bibinfo{author}{\bibfnamefont{C.~H.} \bibnamefont{Keitel}},
  \bibnamefont{and} \bibinfo{author}{\bibfnamefont{R.}~\bibnamefont{Grobe}},
  \bibinfo{journal}{Phys. Rev. A} \textbf{\bibinfo{volume}{89}},
  \bibinfo{pages}{052101} (\bibinfo{year}{2014}),
  \urlprefix\url{https://link.aps.org/doi/10.1103/PhysRevA.89.052101}.

\bibitem[{\citenamefont{Bliokh et~al.}(2017)\citenamefont{Bliokh, Dennis, and
  Nori}}]{Bliohk2017_position}
\bibinfo{author}{\bibfnamefont{K.~Y.} \bibnamefont{Bliokh}},
  \bibinfo{author}{\bibfnamefont{M.~R.} \bibnamefont{Dennis}},
  \bibnamefont{and} \bibinfo{author}{\bibfnamefont{F.}~\bibnamefont{Nori}},
  \bibinfo{journal}{Phys. Rev. A} \textbf{\bibinfo{volume}{96}},
  \bibinfo{pages}{023622} (\bibinfo{year}{2017}),
  \urlprefix\url{https://link.aps.org/doi/10.1103/PhysRevA.96.023622}.

\bibitem[{\citenamefont{Aleksandrov et~al.}(2020)\citenamefont{Aleksandrov,
  Tumakov, Kudlis, Shabaev, and Rosanov}}]{Kudlis}
\bibinfo{author}{\bibfnamefont{I.~A.} \bibnamefont{Aleksandrov}},
  \bibinfo{author}{\bibfnamefont{D.~A.} \bibnamefont{Tumakov}},
  \bibinfo{author}{\bibfnamefont{A.}~\bibnamefont{Kudlis}},
  \bibinfo{author}{\bibfnamefont{V.~M.} \bibnamefont{Shabaev}},
  \bibnamefont{and} \bibinfo{author}{\bibfnamefont{N.~N.}
  \bibnamefont{Rosanov}}, \bibinfo{journal}{Phys. Rev. A}
  \textbf{\bibinfo{volume}{102}}, \bibinfo{pages}{023102}
  (\bibinfo{year}{2020}),
  \urlprefix\url{https://link.aps.org/doi/10.1103/PhysRevA.102.023102}.

\bibitem[{\citenamefont{Sizykh et~al.}(2023)\citenamefont{Sizykh, Chaikovskaia,
  Grosman, Pavlov, and Karlovets}}]{Sizykh}
\bibinfo{author}{\bibfnamefont{G.~K.} \bibnamefont{Sizykh}},
  \bibinfo{author}{\bibfnamefont{A.~D.} \bibnamefont{Chaikovskaia}},
  \bibinfo{author}{\bibfnamefont{D.~V.} \bibnamefont{Grosman}},
  \bibinfo{author}{\bibfnamefont{I.~I.} \bibnamefont{Pavlov}},
  \bibnamefont{and} \bibinfo{author}{\bibfnamefont{D.~V.}
  \bibnamefont{Karlovets}}, \emph{\bibinfo{title}{Transmission of vortex
  electrons through a solenoid}} (\bibinfo{year}{2023}),
  \eprint{arXiv:2306.13161}, \urlprefix\url{https://arxiv.org/abs/2306.13161}.

\bibitem[{\citenamefont{Epp and Guselnikova}(2023)}]{Epp2023}
\bibinfo{author}{\bibfnamefont{V.}~\bibnamefont{Epp}} \bibnamefont{and}
  \bibinfo{author}{\bibfnamefont{U.}~\bibnamefont{Guselnikova}},
  \bibinfo{journal}{Physics Letters A} \textbf{\bibinfo{volume}{469}},
  \bibinfo{pages}{128764} (\bibinfo{year}{2023}), ISSN
  \bibinfo{issn}{0375-9601},
  \urlprefix\url{https://www.sciencedirect.com/science/article/pii/S0375960123001445}.

\bibitem[{\citenamefont{van Kruining et~al.}(2017)\citenamefont{van Kruining,
  Hayrapetyan, and G\"otte}}]{Kruining2017}
\bibinfo{author}{\bibfnamefont{K.}~\bibnamefont{van Kruining}},
  \bibinfo{author}{\bibfnamefont{A.~G.} \bibnamefont{Hayrapetyan}},
  \bibnamefont{and} \bibinfo{author}{\bibfnamefont{J.~B.}
  \bibnamefont{G\"otte}}, \bibinfo{journal}{Phys. Rev. Lett.}
  \textbf{\bibinfo{volume}{119}}, \bibinfo{pages}{030401}
  (\bibinfo{year}{2017}),
  \urlprefix\url{https://link.aps.org/doi/10.1103/PhysRevLett.119.030401}.

\bibitem[{\citenamefont{Gradshteyn and Ryzhik}(2014)}]{Gr-R}
\bibinfo{author}{\bibfnamefont{I.~S.} \bibnamefont{Gradshteyn}}
  \bibnamefont{and} \bibinfo{author}{\bibfnamefont{I.~M.}
  \bibnamefont{Ryzhik}}, \emph{\bibinfo{title}{Table of integrals, series, and
  products}} (\bibinfo{publisher}{Academic press}, \bibinfo{year}{2014}).

\end{thebibliography}

\end{document}